\documentclass[onecolumn, draftcls,journal]{IEEEtran}
\usepackage{amsmath}
\usepackage{amsthm}
\usepackage{comment}
\usepackage{graphicx}
\usepackage{wasysym}
\ifCLASSOPTIONcompsoc
\usepackage[caption=false, font=normalsize, labelfont=sf, textfont=sf]{subfig}
\else
\usepackage[caption=false, font=footnotesize]{subfig}
\fi
\usepackage{resizegather}
\usepackage{color}
\usepackage{cite}
\usepackage{float}
\usepackage{url}
\usepackage[linesnumbered,ruled]{algorithm2e}
\usepackage{epstopdf}
\usepackage{enumitem}
\usepackage{tabularx}
\usepackage{makecell}

\newtheorem{theorem}{\textbf{Theorem}}
\graphicspath{{figures/}}

\newcommand{\edit}[1]{\textcolor{black}{#1}}
\newcommand{\bd}{\mathbf}

\title{Using Battery Storage for Peak Shaving and Frequency Regulation: Joint Optimization for Superlinear Gains}
\author{Yuanyuan Shi, Bolun Xu, Di Wang, Baosen Zhang%
\thanks{Y. Shi, B. Xu and B. Zhang are with the Department of Electrical Engineering at the University of Washington. Emails: {yyshi,xubolun,zhangbao}@uw.edu. These authors are partially supported by the Washington Clean Energy Institute and Microsoft.}%
\thanks{D. Wang is with Microsoft Research. Emails: wangdi@microsoft.com}\vspace{-0.8cm}}
\date{September 2016}
\begin{document}
\vspace{-2cm}
\maketitle
\begin{abstract}
\label{sec:abstract}
We consider using a battery storage system simultaneously for peak shaving and frequency regulation through a joint optimization framework which captures battery degradation, operational constraints and uncertainties in customer load and regulation signals. Under this framework, using real data we show the electricity bill of users can be reduced by up to 12\%. Furthermore, we demonstrate that the saving from joint optimization is often larger than the sum of the optimal savings when the battery is used for the two individual applications. A simple threshold real-time algorithm is proposed and achieves this superlinear gain. Compared to prior works that focused on using battery storage systems for single applications, our results suggest that batteries can achieve much larger economic benefits than previously thought if they jointly provide multiple services.
\end{abstract}

\section{Introduction}
\label{sec:intro}
Battery energy storage systems are becoming increasingly important in power system operations. As the penetration of uncertain and intermittent renewable resources increase, storage systems are critical to the robustness, resiliency, and efficiency of energy systems. For example, studies suggest that 22 GW of energy storage would be needed in California by 2050~\cite{CAISO2014} and the entire United States could require 152 GW of storage~\cite{SolomonEtAl2014}. Much of these capacities are expected to be achieved by distributed storage systems owned by individual consumers~\cite{WorthmannEtAl2015}.

Currently, the two most prominent types of consumers that own significant levels of storage are information technology companies and operators of large buildings. Companies such as Microsoft and Google use battery storages extensively in their data centers as failover to onsite local generation~\cite{Guo2013}. These systems are sized to the capacity of the data center: a 10 MW data center will have a storage system with the power rating of 10 MW with several minutes of energy capacity. In commercial buildings, batteries are used to smooth their load and provide backup services~\cite{WangEtAl2016}. These batteries tend to be slightly smaller, but are still in the 100's of kW/kWh range.

Today, despite their potential to grid services, these battery storage systems are not integrated with the power system. To a storage owner, whether a battery taking part in grid services is predominantly determined by the economic benefits of these services. For example, a data center replaces its batteries every four years or so under normal conditions~\cite{CVD2014}. If the battery participates in the electricity market, batteries may degrade faster and require more frequent replacements. Do the gains from the market justify the additional operational and capital costs?

The question of optimally operating a battery to maximize its economic benefit is a central one and has spurred a substantial body of research. The problems include energy arbitrage, peak shaving, frequency regulation, demand response and others~(e.g. see~\cite{EPRI2013,Shi2016,DunnEtAl2011,eyer2010energy} and the references within). In the past several years, it has been recognized that because of the high capital cost of batteries~\cite{Wasowicz2012}, serving a single application is often difficult to justify their investments~\cite{xi2014stochastic}. In addition, picking a single application does not consider the possibility of multiple revenue streams and may leave ``money on the table''.
Consequently, a recent line of research has started to analyze the co-optimization of batteries for both energy arbitrage and regulation services~\cite{cheng2016co,walawalkar2007economics}.
\begin{figure}
    \centering
    \includegraphics[width=0.8 \columnwidth]{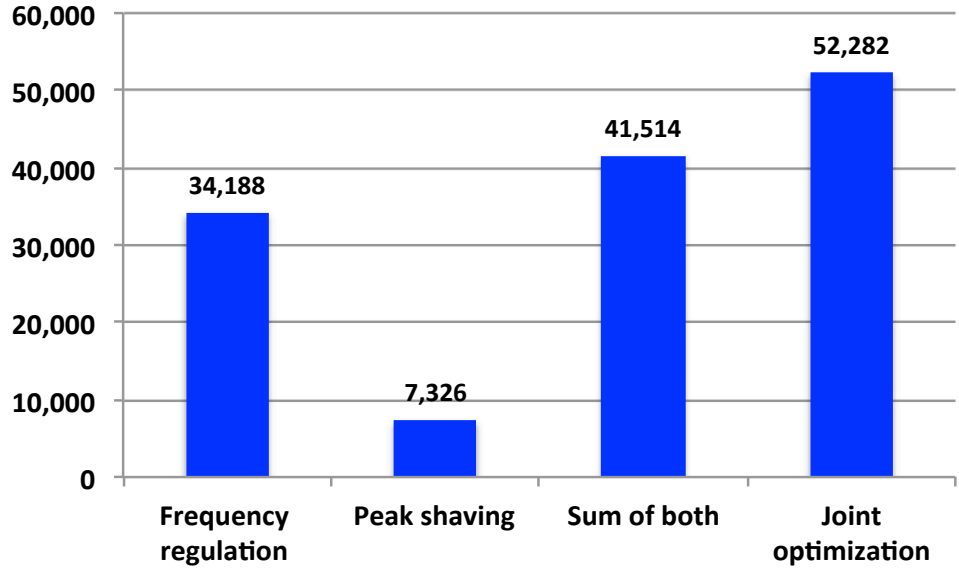}
    \caption{Annual electricity bill savings for a 1MW data center (in PJM control area, total bill of \$488,370) under different battery usage scenarios. Savings from the joint optimization framework proposed in this paper is larger than the sum of savings from frequency regulation service and peak shaving.}
    \label{fig_intro}
\end{figure}

In this paper, we consider the \emph{joint optimization} of using a battery storage system for both \emph{peak shaving} and \emph{frequency regulation} for a commercial customer. Peak shaving can be used to reduce the peak demand charge for these customers and the (fast) frequency regulation is an ideal service to provide for batteries because of their near instantaneous response time. The challenge in combining these two applications lies in their \emph{vastly different timescales}: peak demand charge is calculated every month on a smoothed power consumption profile (e.g. 15-minute averages), while fast frequency regulation requires a decision every 2 to 4 seconds. 

The key observation in our work is that serving different applications over different timescales is economically beneficial to the battery: by exploring the diversity in different applications, we can obtain a so-called \emph{superlinear gain}. An example of the superlinear gain is presented in Fig. 1. It gives the annual electricity bill savings for a 1MW data center under three scenarios, using batteries for frequency regulation service, peak shaving and joint optimization. For joint optimization, we use a simple online threshold algorithm given in Section IV. While for peak shaving and regulation service, the solutions are offline optimal. The super-linear gain arises for reasons that would be explored in depth in the rest of the paper, but briefly speaking, the randomness of frequency regulation signal could contribute to more efficient peak shaving. By exploring the diversity and mutual benefit in different applications, we have this non-linear behavior.

It should be noted that the key function of the existing batteries in data centers is to provide backup capacity. The proposed joint optimization framework is deploying only part of the battery energy capacity while a large portion of the battery energy has been reserved for backup purpose. A more detailed discussion on the division of battery for grid service and backup is provided in Section V-A.

\subsection{Literature Review}
The line of literatures consider co-optimization of storage starts from \cite{walawalkar2007economics}. In \cite{walawalkar2007economics}, the authors analyze the economics of using storage device for both energy arbitrage and frequency regulation service. The work in \cite{white2011using} extended this ``dual-use'' idea by considering plug-in electric vehicles as grid storage resource for peak shaving and frequency regulation. Both works showed that dual-use of storage often leads to higher profits than single applications. However, the aforementioned works mainly rely on heuristic analysis under different price and user patterns without directly using optimization models. The work of~\cite{dowling2017multi} bridged the methodology gap by proposing a systematic co-optimization framework, which could be applied for evaluating different application combinations at different timescales. This framework assumes that all future information is known, so it cannot be extended directly to deal with potential uncertainties from energy and ancillary service markets~(e.g., price, frequency regulation signal, etc.).

To deal with uncertainties, \cite{xi2014stochastic,cheng2016co} formulate the battery co-optimization problem as a stochastic program. In~\cite{xi2014stochastic}, stochastic programming was solved to obtain hourly optimal decisions. The work in \cite{cheng2016co} included applications of different time-scales in its optimization and tackled computational challenges by taking advantage of the problem's nested structure.

\subsection{Our Contributions}
\begin{figure}
    \centering
    \subfloat[Microsoft data center]{%
        \includegraphics[width= 0.4 \columnwidth]{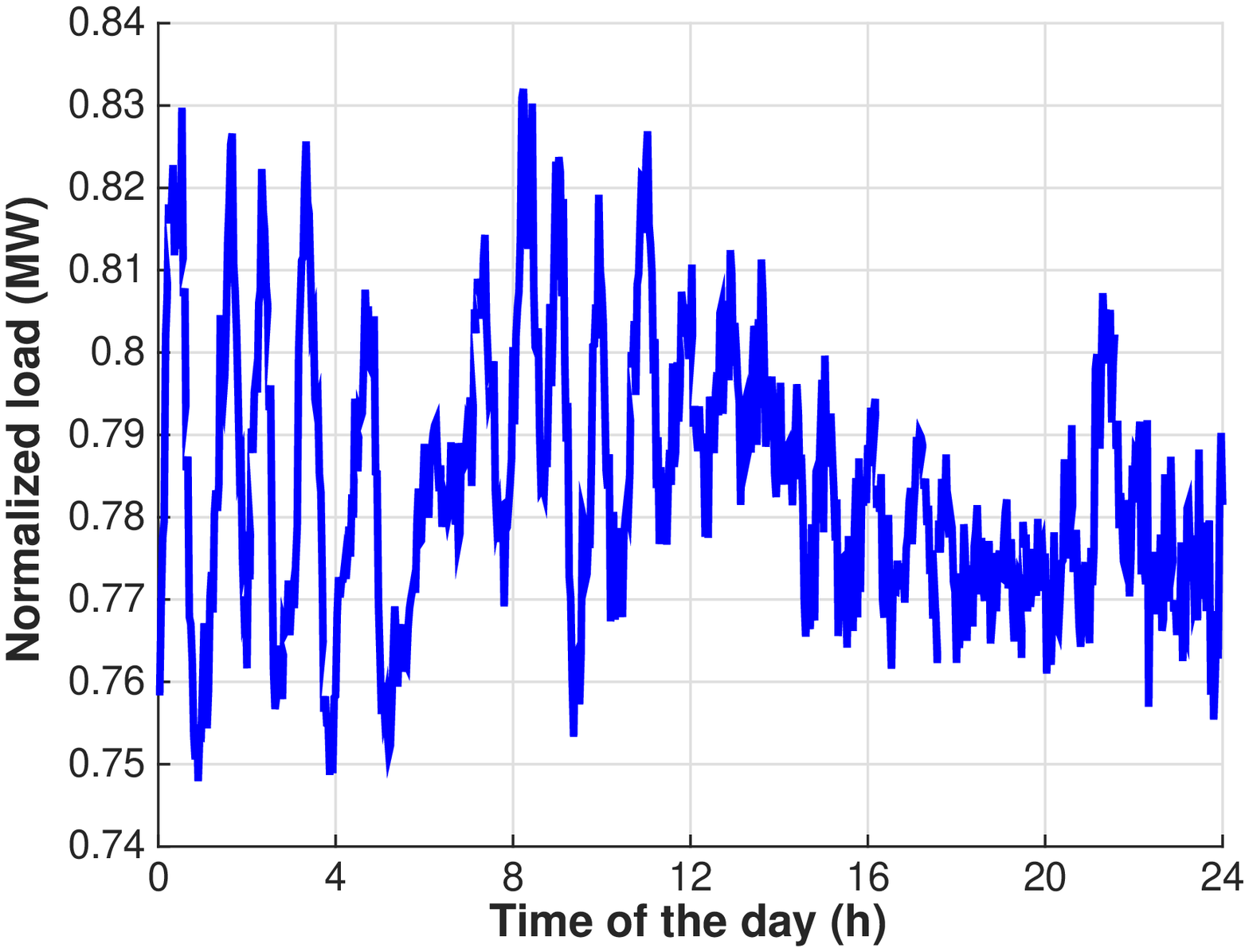}}
    \label{intro:datacenter_load}
    \subfloat[UW EE \& CSE building]{%
        \includegraphics[width=0.4 \columnwidth]{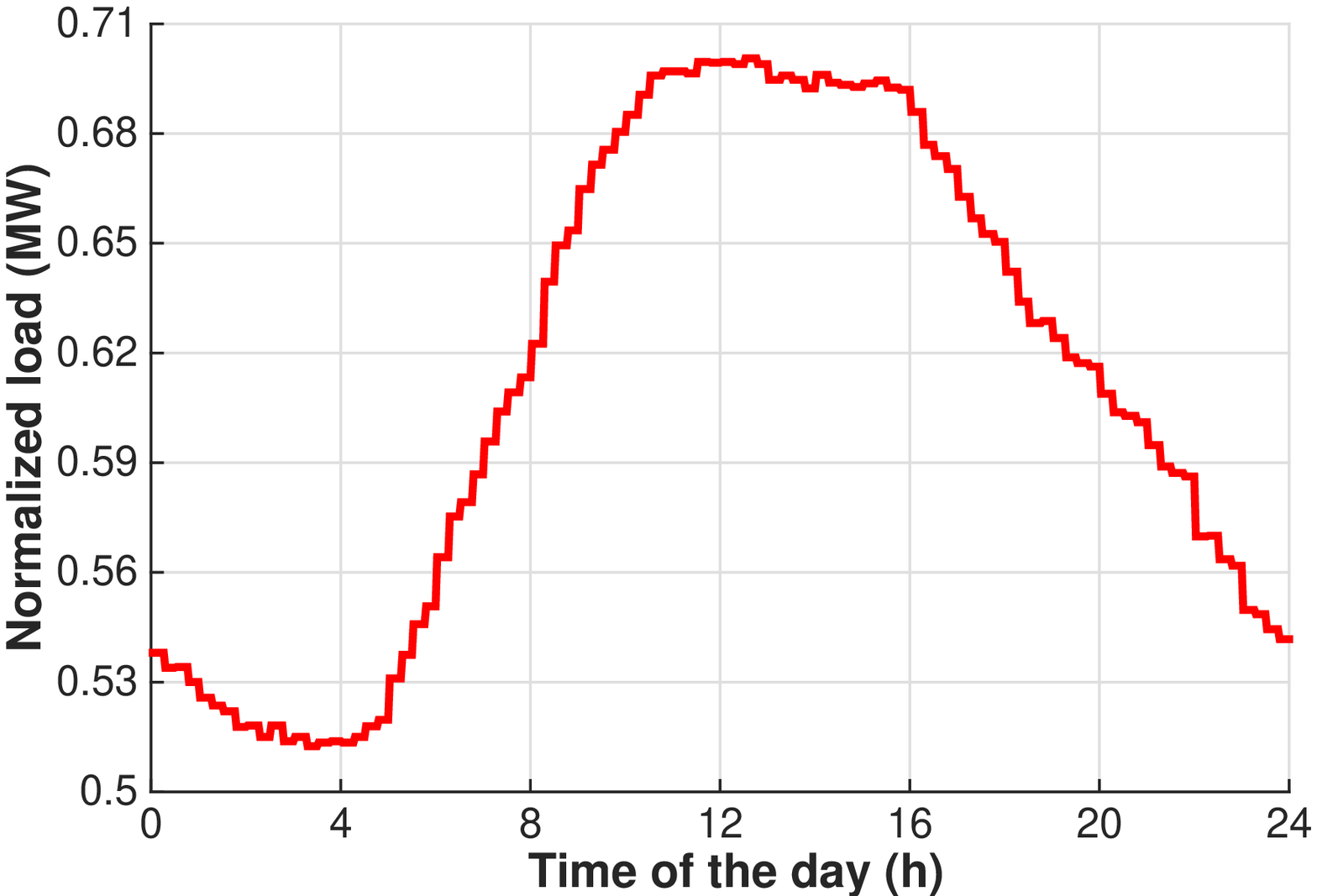}}
    \label{intro:eecs_load}\\
    \caption{Example day load of Microsoft data center and UW EE \& CSE building. Both loads are normalized with respect to their rated power.}
    \label{intro:load}
\end{figure}

Our work is close in spirit to \cite{cheng2016co}, which captures both the future market uncertainties and timescale difference of multiple applications. However, compared with \cite{xi2014stochastic,cheng2016co}, our work contributes in two significant ways:
\begin{itemize}
    \item We propose a joint optimization framework for batteries to perform peak shaving and provide frequency regulation services. This framework accounts for battery degradation, operational constraints, and the uncertainties in both the customer load and regulation signals. All of the previous works, to our knowledge, do not include the operational cost of batteries in their optimization models, which can potentially lead to aggressive charging/discharging responses and severely suboptimal operations~\cite{walawalkar2007economics,white2011using,dowling2017multi,xi2014stochastic,cheng2016co}. Since batteries cycle multiple times a day when used for frequency regulation and peak shaving, the degradation effect plays an important role in determining their operations.
    \item We show that there is a \emph{superlinear gain}: where the revenue from joint optimization is larger than the \emph{sum} of performing the individual applications. We quantify this gain using real world data from two large commercial users: a Microsoft data center and the University of Washington EE \& CSE building. Figure \ref{intro:load} gives an example daily load profile for both cases. The superlinear gain is fundamentally different from previous observations in \cite{walawalkar2007economics, white2011using, cheng2016co} which only compared the revenue from co-optimization with one single application rather than the sum of the applications. The results in \cite{xi2014stochastic} hinted at the relationship between co-optimization revenue and the sum of multiple revenue streams, while mainly focusing on the trade-off between different applications and their ``subadditivity''. The key observation in our work is that batteries can achieve much larger economic benefits than previously thought if they jointly provide multiple services by exploring the diversity of different applications.
\end{itemize}

The rest of the paper is organized as follows. Section \ref{sec:formulation} provides some background on electricity market rules and battery cell degradation. Section \ref{sec:joint_optimization} gives the joint optimization formulation. In Section \ref{sec:control}, we propose a simple threshold control algorithm which achieves the superlinear gain in real time. Section \ref{sec:result} analyzes the superlinear benefits using both real data and synthetic data. Finally, Section \ref{sec:con} concludes the paper and outlines directions for future work.

\section{Problem Formulation}
\label{sec:formulation}
This section provides some basic definitions and the detailed model setup.  We consider a finite time horizon partitioned into $T$ discrete intervals, indexed by $t \in \{1,2,...,T\}$.
This section sets up the overall optimization problem in three steps. First, we explain how the electricity bill is calculated for a large commercial user. Then we focus on two potential applications of using batteries: \emph{peak shaving} and \emph{frequency regulation}. Then we introduce the battery degradation model considered in this paper.
\subsection{Electricity Bill of Commercial Users}
We consider commercial consumers whose electricity bill consists of two parts: energy charge and peak demand charge. Let $s(t)$ be the power consumption at at $t$ and $t_s$ be the size of a time step. Then the energy charge is given by:
\begin{equation}
J^{elec} = \lambda_{elec} \sum_{t=1}^{T} s(t) \cdot t_s,
\label{equ:bill_elec}
\end{equation}
where $\lambda_{elec}$ is the price of energy with a unit of $\$/MWh$. The peak demand charge is based on the maximum power consumption. In practice, this charge is calculated from a running average of power consumption over 15 or 30 minutes. Let $\bar{s}(t)$ denote the smoothed demand and then the peak demand charge can be written as,
\begin{equation}
    J^{peak} = \lambda_{peak} \max_{t = 1,2,...,T} [\bar{s}(t)].
\end{equation}
For the rest of the paper, the time step size $t_s$ is absorbed into the price coefficients for simplicity. Hence, the total electricity bill for a commercial user over time $T$ is,
\begin{equation}\label{equ:bill_total}
    J=\lambda_{elec} \sum_{t=1}^{T} s(t)+ \lambda_{peak} \max_{t = 1,2,...,T} [\bar{s}(t))]\,,
\end{equation}
This cost function is convex in $s(t)$ since it is a linear combination of linear functions and a piece-wise max function. In this paper, we investigate how to reduce the total cost \eqref{equ:bill_total} by using battery energy storage (BES). Specifically, we consider two applications, peak shaving and frequency regulation.

\subsection{Peak Shaving}
The peak demand charge of commercial users could be as large as their energy cost. Therefore smoothing or flattening peak demand represents an important method of reducing their electrical bills. A myriad of methods for peak shaving have been proposed in the literature, e.g., using energy storage \cite{Sigrist2013}, load shifting and balancing \cite{Caprino2014}. Here we focus on using batteries. Batteries can discharge energy when demand is high and charge in other times to smooth user's consumption profiles. Let $b(t)$ denote the power injected by the battery, with the convention that $b(t)>0$ represents discharging and $b(t)<0$ represents charging. Then $s(t)-b(t)$ is the actual power draw from the grid. Let $\bd b=[b(1) \; \dots \; b(T)]$ be the vector of battery actions. The total electricity bill becomes
\begin{align}
    J^{P} &= \lambda_{elec} \sum_{t=1}^{T} {[s(t)-b(t)]} + f(\bd b) \nonumber \\
    & + \lambda_{peak} \underset{t=1...T}{\text{max}}[(\bar{s}(t)-\bar{b}(t))]
\end{align}
where  $\bar{b}(t)$ is the averaged power injection of the battery and $f(\bd b)$ models the degradation effect of using the battery. We note that battery degradation costs maybe somewhat complicated functions of the entire profile. More details about the battery cost model will be introduced in Section~\ref{sec:bat}.

\subsection{Frequency Regulation Service}
\begin{figure}
    \centering
    \includegraphics[width=0.4 \columnwidth]{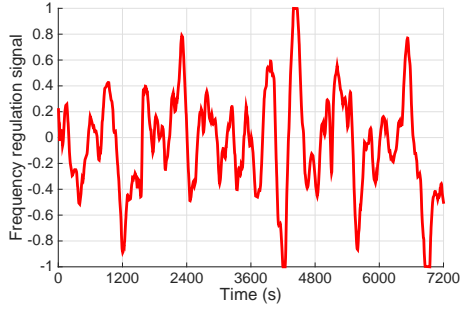}
    \caption{PJM fast frequency regulation signal for 2 hours\cite{pjm}.}
    \label{fig:case_reg}
\end{figure}

Besides doing peak shaving, commercial users could earn revenue by providing grid services. In this paper, we consider using batteries owned by these users to participate in the frequency regulation market. In particular, we adopt a simplified version of the PJM frequency regulation market~\cite{xu2016pesgm}. Fig. \ref{fig:case_reg} gives an example of the PJM fast frequency regulation (RegD) signal for 2 hours. Compared with traditional frequency regulation signals, it has a much faster ramping rate and is designed to have a zero-mean within a certain time interval, which is well aligned with the characteristics of batteries.

For providing frequency regulation service, the grid operator pays a per-MW option fee $\lambda_c$ to a resource with stand-by power capacity $C$ for each hour. While during the frequency regulation procurement period, the resource is subjected to a per-MWh regulation mismatch penalty ($\lambda_{mis}$) for the absolute error between the instructed dispatch and the resource's actual response. Let $r(t)$ be the normalized RegD signal, and the revenue from providing frequency regulation service over time $T$ is:
\begin{equation}
R = \lambda_c C \cdot T - \lambda_{mis} \sum_{t=1}^{T}{|b(t)-Cr(t)|} - f(\bd b),
\end{equation}
where $f(\bd b)$ is again the operating cost of the battery.

\subsection{Battery Cell Degradation} \label{sec:bat}
A key factor in the operational planning of battery energy storage (BES) is its operating cost, a majority of which stems from the degradation of battery cells subjected to repeated charge/discharge cycles. Different batteries exhibit different degradation behaviors, and their understanding and characterization is a major area of study (see \cite{Miguel2011} and references within). In this paper, we focus on lithium-ion batteries, which are one of the most popular batteries used in practice today.

Modeling battery degradation is a challenging task and no single model can be used for all types of chemistry. We do not attempt to propose a single detailed model in this paper; instead, some general features of battery degradation are captured and included in our optimization framework.
 \begin{figure}
     \centering
     \includegraphics[width= 3.5in, height = 1.6in]{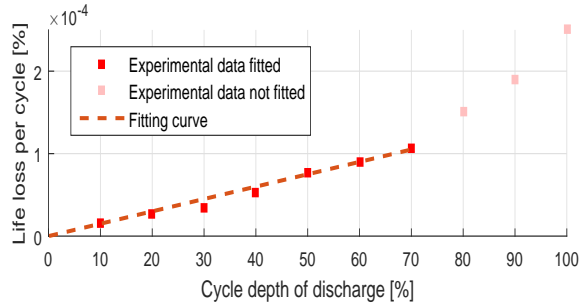}
     \caption{LMO battery cycle life curve VS. Cycle numbers}
     \label{fig_lmo}
 \end{figure}

 For instance, the authors in~\cite{Xu2016degradation} showed that the capacity of the Lithium Manganese Oxide (LMO) batteries are sensitive to both the number of cycles as well as the cycle depth of discharge (DoD). The data in Fig. \ref{fig_lmo} comes from a LMO cycle life test data in~\cite{Xu2016degradation}. It shows that, if we limit the battery operation within certain DoD region to avoid the overcharge and over-discharge effects, there is a constant marginal cost for the cycle depth increase. In addition, cycle DoD is a linear function of the amount of battery charging/discharging,
\begin{align}
DoD = \sum_{t\in T_i} b(t)t_s,
\end{align}
where $T_{i}$ denotes a set of timesteps belongs to the same battery degradation cycle. Thus, we could assign a constant marginal cost for the amount of battery charging/discharging:
$f(\bd b) \propto \lambda_b |b(t)|,$
where $\lambda_{b}$ is the linearized battery degradation cost co-efficient. To get $\lambda_{b}$, we normalize battery lifetime into the amount of energy a battery cell can process before reaching end-of-life, and prorate the battery cell cost into a per-MWh cost with respect to the charged and discharged energy.
\begin{equation}
\lambda_b = \frac{\lambda_{cell} \cdot 10^{6}}{2N \cdot (SoC_{max} - SoC_{min})}
\label{eq: lambda_b}
\end{equation}
In the above formula, $\lambda_{cell}$ is the battery cell price (\$/Wh), N is the number of cycles that the battery could be operated within SoC limit $[SoC_{min}, SoC_{max}]$.

\edit{For Lithium-ion batteries based on other chemistries such as lithium iron phosphate ($LiFePO_4$), the loss of life per cycle is interpreted as a function of cycle numbers rather than the amount of energy charged or discharged~\cite{Guo2013}.
This type of degradation model works well for long timescale applications considered in~\cite{Guo2013}. However, we consider battery for fast frequency regulation service, where the direction of $b(t)$ changes fairly quickly. If one interprets the degradation cost as a function of the charging/discharging direction change times, the model would be too aggressive. In fact, this model suggests that a battery may die in a matter of days. Therefore, we adopt a cost that is proportional to the battery power. We limit the battery operation within certain DoD range (70\%), and assign a constant marginal cost for battery energy charging and discharging.}

\section{Joint optimization framework}
\label{sec:joint_optimization}
\subsection{The Joint Optimization Model}
In this paper, we consider using a battery to provide frequency regulation service and peak shaving simultaneously, thus to boost the economic benefits. The stochastic joint optimization problem is given in (\ref{Sec3:P4:1}), which captures both the uncertainty of future demand $s(t)$ and the uncertainty of future frequency regulation signals $r(t)$.

\begin{small}
 \allowdisplaybreaks\begin{subequations}\label{Sec3:P4:1}
 	\begin{align}
 	J^{joint} = & \min_{C, b^{ch}(t), b^{dc}(t), y(t)} \; \lambda_{elec} \sum_{t=1}^{T} {E_{\bd s}\left[s(t)- b(t)\right]} \nonumber\\
 	& + \lambda_{peak} \underset{t=1...T}{\text{max}} E_{\bd s}\left[\bar{s(t)}-  \bar{b(t)}\right]+\sum_{t=1}^{T} {f(b(t))} \nonumber\\
 	&  \!-\!E_{\bd r, \!\bd s} \!\left[\lambda_c T \!\cdot C \!- \!\lambda_{mis} \sum_{t=1}^{T}{|-\!s(t)\!+ b(t)\!+y(t)\!-\!Cr(t)|}\right] \label{co:obj}\\
 	\text{s.t.\ }  & b(t) = b^{dc}(t)-b^{ch}(t)\,, \\
 	&  \ C \geq 0\,,  \label{co:pc}\\
 	& SoC_{min} \!\leq \!\frac{SoC_{ini} \!+ \!\sum_{\tau\!=\!1}^{t}\!\left[\!b^{ch}\!(\tau\!)\eta_{c}\!-\!\frac{b^{dc}\!(\tau\!)}{\eta_{d}}\right]\!t_s}{E} \!\leq\!SoC_{max} \label{co:soc}\\
 	& 0\ \leq b^{ch}(t) \leq P^{max}\,,\label{co:charging}\\
 	& 0\ \leq b^{dc}(t) \leq P^{max}\,.\label{co:discharging}
 	\end{align}
 \end{subequations}
\end{small}

The objective function (\ref{co:obj}) minimizes the total electricity cost of a commercial user for the next day, including the energy cost, peak demand charge, battery degradation cost and frequency regulation service revenue. The optimization variables are frequency regulation capacity $C$, battery charging/discharging power $b^{ch}(t)$, $b^{dc}(t)$ and frequency regulation load baseline $y(t)$. Participants in frequency regulation market should report a baseline $y(t)$ to the grid operator ahead of their service time \cite{xu2016pesgm}. For a commercial user, the baseline $y(t)$ is its load forecasting. Constraint (\ref{co:pc}) guarantees a non-negative frequency regulation capacity bidding. \eqref{co:soc}, \eqref{co:charging} and \eqref{co:discharging} represent the battery SoC limit and power limits.

\subsection{Benchmark}
To show the gain of joint optimization, we describe two benchmark problems: the offline (deterministic) peak shaving problem and the offline (deterministic) frequency regulation service problem. In these benchmarks, we assume complete knowledge of the future. In essence, the benchmarks here represent the best possible performance of any algorithms that solve these problems individually.

The offline peak shaving problem is:
\begin{small}
\begin{subequations}\label{Sec3:ps}
 	\begin{align}
 	J^{p} = \min_{b(t)} \; &\lambda_{elec} \sum_{t=1}^{T} {[s(t)-b(t)]} + \lambda_{peak} \underset{t=1...T}{\text{max}}[\bar{s(t)}-\bar{b(t)}] \nonumber \\
 	&  +  \sum_{t=1}^{T} f(b(t)) \label{ps:obj}\\
 	\text{s.t.\ }  & b(t) = b^{dc}(t)-b^{ch}(t)\,, \\
 	& SoC_{min} \!\leq \!\frac{SoC_{ini} \!+ \!\sum_{\tau\!=\!1}^{t}\!\left[\!b^{ch}\!(\tau\!)\eta_{c}\!-\!\frac{b^{dc}\!(\tau\!)}{\eta_{d}}\right]\!t_s}{E} \!\leq\!SoC_{max} \label{ps:soc}\\
 	& 0\ \leq b^{ch}(t) \leq P^{max}\,,\label{ps:charging}\\
 	& 0\ \leq b^{dc}(t) \leq P^{max}\,.\label{ps:discharging}
 	\end{align}
\end{subequations}
\end{small}

The above problem is convex in terms of $\bd b$. We solve it and denote the optimal bill value as $J^p$. The offline frequency regulation problem is:

\begin{small}
	\begin{subequations}\label{Sec3:reg}
 	\begin{align}
 	R^{*} = \max_{C, b(t)} \;
 	&  \lambda_c T \cdot C - \lambda_{mis} \sum_{t=1}^{T}{|b(t)-Cr(t)|} - \sum_{t=1}^{T} f(b(t)) \label{reg:obj}\\
 	\text{s.t.\ }  & b(t) = b^{dc}(t)-b^{ch}(t)\,, \\
 	& C \geq 0 \,, \label{reg:capacity}\\
 	& SoC_{min} \!\leq \!\frac{\!SoC_{ini} \!+ \!\sum_{\tau\!=\!1}^{t}\!\left[\!b^{ch}\!(\tau\!)\eta_{c}\!-\!\frac{b^{dc}\!(\tau\!)}{\eta_{d}}\right]\!t_s}{E} \!\leq\!SoC_{max} \label{reg:soc}\\
 	& 0\ \leq b^{ch}(t) \leq P^{max}\,,\label{reg:charging}\\
 	& 0\ \leq b^{dc}(t) \leq P^{max}\,.\label{reg:discharging}
 	\end{align}
 \end{subequations}
\end{small}

The above regulation revenue maximization problem does not consider the effect of providing frequency regulation service on electricity bills. Recall that frequency regulation is a service managed by grid operators, while as an end consumer, the commercial user's electricity supply contracts with the utility is unchanged, thus the user still subjects to the energy and peak demand charge. Therefore, the overall electricity bill $J^r$ is,
\begin{align}\label{Sec3:P2:3}
 J^{r} &= \lambda_{elec} \sum_{t=1}^{T} {[s(t)-b^{r}(t)]} \nonumber\\
 & + \lambda_{peak}  \underset{t=1...T}{\text{max}}[\bar{s(t)}-\bar{b}^{r}(t)] -{R^{*}}\,,
\end{align}
where $b^{r}(t)$ is the optimal battery responce for frequency regulation service and $R^{*}$ the optimal service revenue.

\edit{Both of the benchmark problems are convex because all of the constraints are linear and objectives are an addition of convex functions (pointwise maximum is a convex function). To solve these problems, we use the CVX package for Matlab~\cite{cvx}, a generic package for solving convex problems. We used a 2.5 GHz Intel Core i7 Macbook with 16 GB memory.  The problem size can be fairly large, since the time resolution is 4s. But even for an 8-hour horizon, the problem can be solved in about 10 minutes.}

\subsection{The Superlinear Gain}
\label{sec:superlinear}
Our results highlight that a \emph{superlinear} gain can often be obtained: the saving from the \emph{stochastic} joint optimization can be larger than the \emph{sum} of two benchmark optima. In mathematical form, superlinear gain denotes the following phenomenon, which often holds in practice,
\begin{equation}\label{eq:superlinear}
J-J^{joint} > (J-J^{r})+(J-J^{p})\,,
\end{equation}
where the left side of Eq. (\ref{eq:superlinear}) is the saving from joint optimization, and the right side represents the sum of savings from two benchmark problems. The key observation in such case is the ``super-additivity''. The revenue from co-optimization of multiple applications is not only higher than the revenue from any single application (which may be obvious), but also higher than the sum of revenues from all individual applications.

\edit{We provide an example to demonstrate the superlinear gain. Table \ref{table_datacenter} gives the daily electricity bill under four scenarios for a 1MW data center: the original bill (batteries are left idle), using battery only for frequency regulation service, using battery only for peak shaving, and using battery for both services (detailed algorithm discussed in Section \ref{sec:control}). The bill savings are highlighted, from which we observed that saving from joint optimization is larger than the sum of each individual applications. The load curve and frequency regulation signals for that day are given in Fig. 2a and Fig. \ref{fig:case_reg}.}
\begin{table*}
	\renewcommand{\arraystretch}{1.5}
	\centering
	\caption{\edit{One day electricity bill for Microsoft data center under four scenarios: original bill, bills after providing frequency regulation service, peak shaving, and joint optimization. Bill savings and saving ratios are bolded.}}
	\begin{tabular}{lllllll}
		\hline
		\hline
		& Total bill (\$)  & Bill saving (\$)  & Energy charge (\$) & Peak charge (\$)  & Battery cost (\$) & Regulation rev (\$) \\
		\hline
		Original 			&1345.7 & \bf{0}       & 884.2 & 461.5  &0 & 0\\
		Frequency regulation		  &1254.6 & \bf{91.1 (6.77 \%)}& 884.2 & 528.7 & 123.1 & 301.4\\
		Peak shaving	&1331.9 & \bf{23.8 (1.76\%)}  & 884.2 & 424.8 & 12.9 & 0 \\
		Joint optimization  &1194.5 & \bf{151.2 (11.24\%)}   & 884.2 & 465.8 & 117.2 & 272.7 \\
		\hline
		\hline
	\end{tabular}
	\label{table_datacenter}
\end{table*}

\section{Online Battery Control}
\label{sec:control}
\begin{figure}
	\centering
	\includegraphics[width=2.5in]{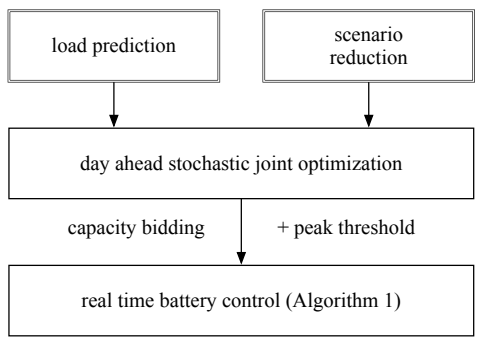}
	\caption{Work flow of the proposed battery control method. We use load prediction and scenario reduction to solve the day-ahead stochastic optimization problem. Then we feed in the capacity bidding and peak threshold for real-time control.}
	\label{fig:joint_control}
\end{figure}

The previous section demonstrates that battery joint optimization could have superlinear gains. In this section, we propose an online battery control algorithm for joint optimization. The challenge in combining peak shaving and frequency regulation service together lies in their vastly different timescales. To deal with the timescale difference, we divided the optimization problem into two stages, 1) day-ahead decision on peak shaving threshold and frequency regulation capacity bidding; 2)  real-time control of battery charging/discharging. Fig. \ref{fig:joint_control} summarizes the workflow of the overall control algorithm. In this section, we first introduce the load prediction and scenario reduction method for solving the day-ahead optimization problem, and then a real-time battery operation algorithm is presented.
\subsection{Load Prediction and Scenario Generation}
\label{sec:load_scenario}
\edit{We use a multiple linear regression (MLR) model \cite{hong2010} for day-ahead load prediction, which is simple, easy to implement in commercial user's site, yet achieves high prediction accuracy. Details of the load prediction algorithm are given in Appendix A. We used the 10-fold cross validation method to evaluate the MLR load prediction model, and the resulting mean absolute percentage error (MAPE) is $3.7\%$ for Microsoft data center load and $2.3\%$ for University of Washington EE \& CSE building. MAPE is a measure of prediction accuracy, which is calculated by averaging the absolute deviation between real value and prediction divided by the actual value.}

\edit{To deal with the uncertainty of future regulation signal, a scenario-based method is implemented. Here, we use one-year historical data to empirically model the distribution of regulation signals. Each daily realization of the regulation signal is called a ``scenario'', and thus we obtain 365 scenarios. We applied the forward scenario reduction algorithm in \cite{wang2015} to select the best subset of scenarios. We set the number of selected scenarios as 10, which strives for a balance between performance and computational complexity by simulation results. Therefore, we have a set of 10 scenarios for frequency regulation signals denoted as $\Omega$ and each scenario associated with a realization probability $\omega_{i}$, which in total compose the uncertainty set of frequency regulation signal.}

We solve the stochastic problem in (\ref{Sec3:P4:1}) using the load prediction $\hat{s}(t)$ and contructed regulation signal uncertainty sets $\Omega$. Define the optimal battery response and frequency regulation capacity as $b^{*}(t)$ and $C^{*}$, then the optimal peak shaving threshold $U^{*}$ is,
\begin{equation}
U^{*} = \underset{t=1...T}{\text{max}} \big[\overline{\hat{s}}(t)-\bar{b}^{*}(t)\big]\,,
\end{equation}

\subsection{Real-time Control for Battery Charging/Discharging}
\label{sec:sim_control}
Section \ref{sec:load_scenario} describes how to make day-ahead decisions on capacity bidding and peak threshold. Here we introduce a simple real-time battery control algorithm for joint optimization. It is computationally efficient, which only requires the measurement of battery's real-time state of charge (SoC) and achieves near-optimal performance compared with the offline optima with perfect foresight. \edit{Although more sophisticated methods such as model predictive control~\cite{qin2016} or dynamic programming~\cite{cheng2016co, xi2014stochastic} have been proposed in the previous literature, they are not needed in this case given the near-optimal performance and high computational efficiency of the proposed online control method.} 

The intuition for the real-time joint optimization control algorithm comes from the optimal battery control algorithm for frequency regulation service. Recall the benchmark frequency regulation problem with objective \eqref{reg:obj}: under linear battery cost model, given a fixed capacity bidding $C$, we have a simple yet optimal real-time battery control method. Theorem \ref{theorem1} describes the optimal control algorithm for batteries providing frequency regulation service.

\begin{theorem}
	Assume $\lambda_b < \lambda_{mis}$.\footnote{Otherwise the battery would not be used at all.} If the marginal battery charging/discharging cost is constant within the operation region, that is, $f(b(t)) = \lambda_b |b(t)|$. For a given capacity $C$, the optimal battery response $b^{*}(t)$ for providing frequency regulation service is:
\edit{\begin{itemize}
	\item $\min\{Cr(t), P^{max}, \frac{\eta_d [soc(t)-SoC_{min}]E}{t_s}\}$, if $r(t)\geq 0$
	\vspace{6pt}
	\item $\max\{Cr(t),-P^{max} ,\frac{[SoC(t)-SoC_{max}] E}{\eta_c t_s}\}$,  if $r(t) < 0$
\end{itemize}}
where $SoC(t)$ of the battery state of charge at the beginning of time step $t$.

	\label{theorem1}
\end{theorem}

The proof of Theorem \ref{theorem1} is given in the Appendix C. As the theorem shows, when the marginal operation cost for battery charging/discharging is constant, the optimal battery control policy is a simple threshold policy. Following Theorem \ref{theorem1}, we propose a real-time control algorithm for joint optimization in Algorithm \ref{algorithm:joint}.
\begin{table}
	\centering
	\caption{\edit{Comparison between the simple threshold algorithm and the offline solution with full information}}
	\begin{tabular}{ccc}
		\hline
		\hline
		& \thead{Data center\\} & \thead{UW  EE \& CSE\\} \\
		\hline
		\thead{Total days simulated\\}			&183 & 365\\
		\thead{Average original daily bill (\$)\\} & 1338.1 & 985.3\\
		\thead{Average joint optimization bill\\ with perfect insight (\$)} & 1184.3 & 857.2\\
		\thead{Average joint optimization bill \\with simple online control (\$)} & 1194.9 & 863.6\\
		\hline
		\hline
	\end{tabular}
	\label{table:simcontrol}
\end{table}
\edit{Table \ref{table:simcontrol} gives a comprehensive comparison between the simple online control algorithm and the offline optima with perfect foresight based on half year of simulation results of Microsoft data center and one-year data of UW EE \& CSE building. Both results show that by implementing the simple threshold algorithm, \emph{we can achieve near-optimal performance compared with the perfect foresight case}.}
\begin{algorithm}
    \SetKwInOut{Input}{Input}
    \SetKwInOut{Output}{Initialize}
    \SetKwComment{Comment}{$\triangleright$\ }{}
    \Input{$C^{*}$, $U^{*}$, $P^{max}$, $E$}
    \Output{$t=0$, $SoC(1)=SoC_{init}$}
    \For {$t = 1 \to T$}{
        Receive regulation signal, $r(t)$.

        Locate the current time $t$ in its corresponding peak calculated window, $[\tau_o, \tau_e]$.

        Calculate peak demand value of the current period,
        $$U = \frac{\sum_{\tau=\tau_o}^{t} [s(\tau)-b(\tau)]}{(t-\tau_o)}\,,$$

		 \tcc{Simple control}
         \uIf{$U \leq U^{*}$}{
                $b(t) = C^{*} \cdot r(t)$\;
           }
            \Else{
                $b(t) = C^{*} \cdot r(t) + (U-U^{*})$ \;
            }

        \tcc{Power and SoC thresholds}
        \uIf{$b(t) \geq 0$}{
                $b(t) = \min\{b(t), P^{max}, \frac{\eta_d [soc(t)-SoC_{min}]E}{t_s}\}$
                $b^{dc}(t) =  b(t), b^{ch}(t) = 0$\tcp*{discharge}
            }
            \Else{
                $b(t) = \max\{b(t), -P^{max}, \frac{[SoC(t)-SoC_{max}] E}{\eta_c t_s}\}$ $b^{ch}(t) =  |b(t)|, b^{dc}(t) = 0$ \tcp*{charge}
            }

        Update the battery SoC status,
        
        $$soc(t+1) = soc(t) + \frac{[b^{ch}(\tau)\eta_{c}-{b^{dc}(\tau)}/{\eta_{d}}]t_s}{E}\,,$$
        Proceed to the next control step: $t \leftarrow t+1$
    }
    \caption{Real-time control for joint optimization}
    \label{algorithm:joint}
\end{algorithm}

\section{Simulation Results}
\label{sec:result}
We provide a case study using half year power consumption data from Microsoft data center and one year data from University of Washington EE \& CSE building. The frequency regulation signal is from PJM fast frequency regulation market\cite{pjm}, where the considered Microsoft data center locates. We implement the simple threshold control algorithm in Section \ref{sec:control} for battery joint optimization. Simulation results demonstrate that over 80\% of time, we will have the superlinear benefits by joint optimization.
\subsection{Parameter Setup}
Assume that the battery optimization horizon is 1 day and the time granularity of t is $4s$, so that $T=4320$. The electricity price is $47 \$/MWh$ and peak demand charge is $12 \$/kW$ per month. For frequency regulation service, suppose the capacity payment is $50\$/MWh$ and set mismatch penalty to guarantee at least $80\%$ performance score \cite{xu2016pesgm}. The BES for optimization is Lithium Manganese Oxide (LMO) battery, with high power capacity and low energy capacity. Within the SoC operation region $SoC_{min} = 0.2$ and $SoC_{max} = 0.8$, LMO battery has a constant marginal degradation cost with regard to how much energy is charged and discharged.

\edit{In this work, we consider using existing batteries in commercial users, e.g., the backup batteries in data centers, to participate in power market and reduce users' electricity bills. The key function of these batteries for users is to provide backup capabilities and the proposed joint optimization framework is deploying only part of the battery energy capacity. We assume the overall battery has a 1MW power capacity and 15 minutes energy capacity, which is a typical size of an industrial-scale grid-tied battery. Then, different portions of the total energy capacity are considered for grid service, 3 minutes, 5 minutes and 10 minutes respectively. The results are presented in Table~\ref{tab:comp2}, where the metrics for comparison under different scenarios are the annual bill savings and battery life expectancy.}
\begin{table}
	\renewcommand{\arraystretch}{1.5}
	\centering
	\caption{\edit{Comparison between different portions of battery energy capacity used for joint optimization.}}
	\begin{tabular}{lcc}
		\hline
		\hline
		Joint optimization capacity &  Annual bill saving  & Life expectancy (Year) \\
		\hline
		3 minutes &  \$52,282 &  3.58 \\
		5 minutes & \$80,547 & 1.34 \\
		10 minutes & \$105,140 & 0.86 \\
		\hline
		\hline
	\end{tabular}
	\label{tab:comp2}
\end{table}
\edit{An aggressive user may try to replace their battery in a yearly basis for the largest bill reduction. More likely, for a building or a data center, a 3 year cycle is preferred. In fact, most data centers already replace their batteries every 3 to 4 years for reliability reasons \cite{CVD2014,wang2012energy}, so using 3 minutes of the battery capacity for grid services would lead to considerable gains without any additional burdens. Of course, the remaining portion of battery energy storage is reserved for emergency backup.}

Therefore we assume for joint optimization usage, the battery power rating $P^{max}$ is $1MW$, energy capacity $E$ is 3 minute,  and battery cell price is $0.5 \$/Wh$. Accoring to Fig. \ref{fig_lmo}, the LMO battery can be operated for $N = 10,000$ cycles when the average cycle DoD is $60\%$. Using Eq. (\ref{eq: lambda_b}), we calculate the battery degradation cost as $83\$/MWh$.

In order to evaluate the performance of the proposed battery joint optimization algorithm, we compare the savings from joint optimization with the sum of savings from benchmark peak shaving and frequency regulation service. A criteria $q$ (joint optimization saving ratio) is defined as below,
\begin{equation}\label{equ:superlratio}
q = \frac{(J-J^{joint})-[(J-J^{r})+(J-J^{p})]}{J}\,,
\end{equation}
which describes the percentile of superlinear saving compared to the original bill.

\subsection{Results for Synthetic Load: Peak Shape and Superlinear Gain}
\edit{In the previous sections (Figure 1 and Table II), we observed that by doing battery joint optimization, we have the superlinear gain. One natural question may come up, why we have the superlinear gain? Before we dive into more simulations on real data, we pick a simple rectangle peak where the base load is 0.5 MW, peak load is 1MW in this section for analysis, in order to better understand the conditions that lead to superlinear gains. We change the duration of peak from 3 minutes (a sharp peak) to 15 minutes (a flat peak) in order to study the effect of peak shape on the probability of superlinear gain.}
\begin{figure}
	\centering
	\subfloat[Frequency regulation only]{%
		\includegraphics[width=0.5\linewidth]{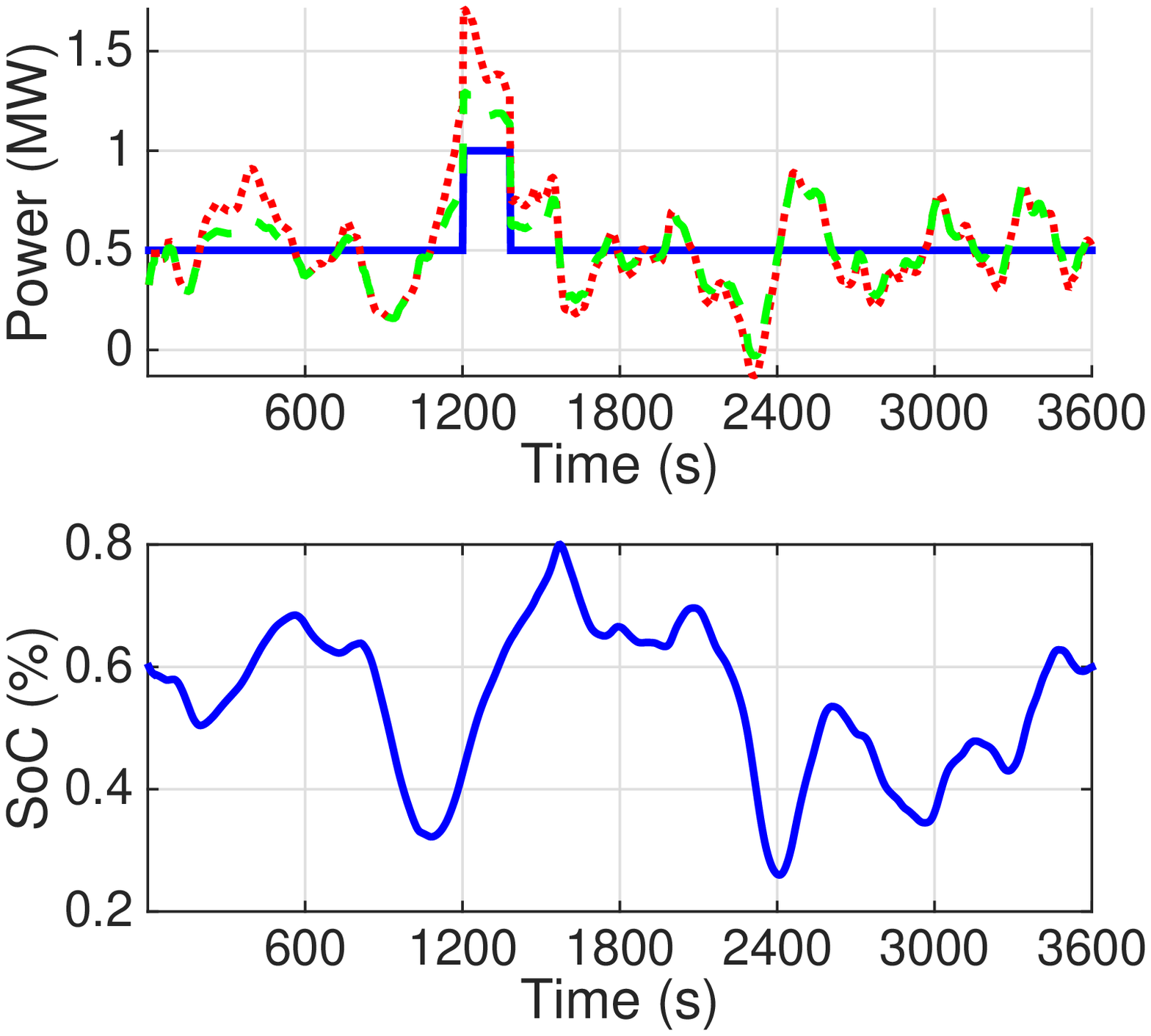}}
	\label{Sec4:P2:1}\hfill
	\subfloat[Peak shaving only]{%
		\includegraphics[width=0.5\linewidth]{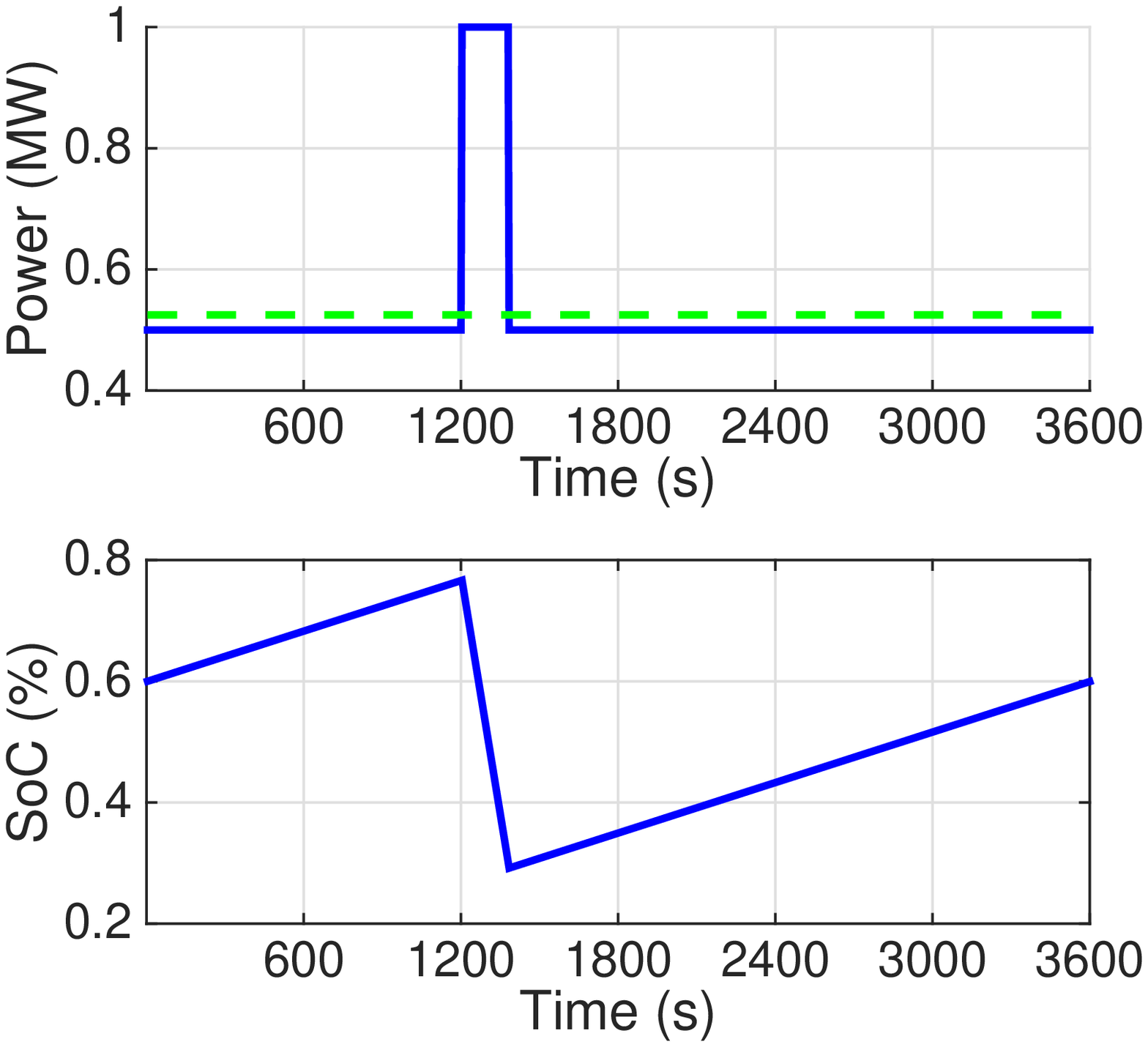}}
	\label{Sec4:P2:2}\\
	\subfloat[Joint optimization]{%
		\includegraphics[width=0.5\linewidth]{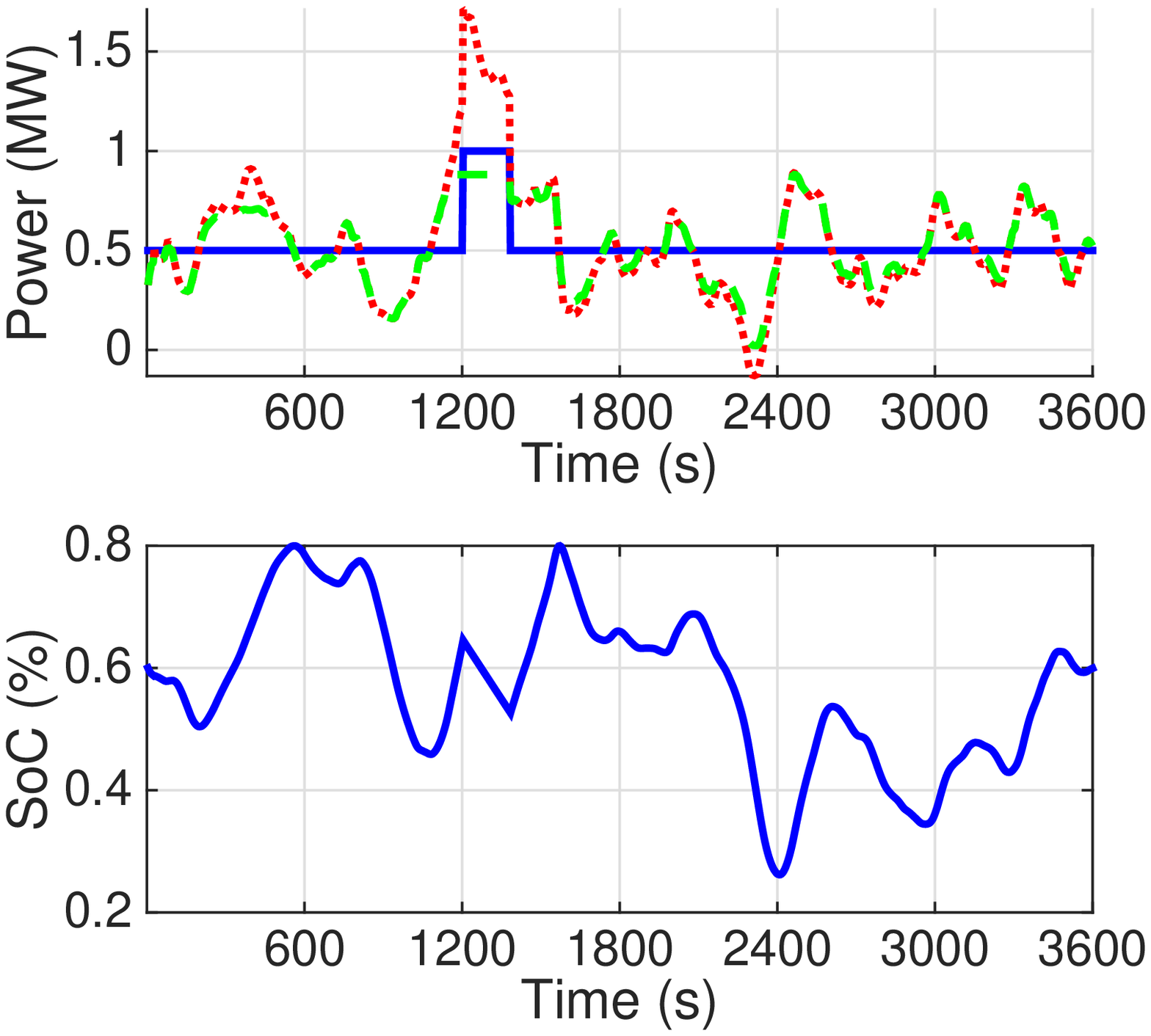}}
	\label{Sec4:P2:3}\hfill
	\subfloat[Bills comparison]{%
		\includegraphics[width=0.5\linewidth]{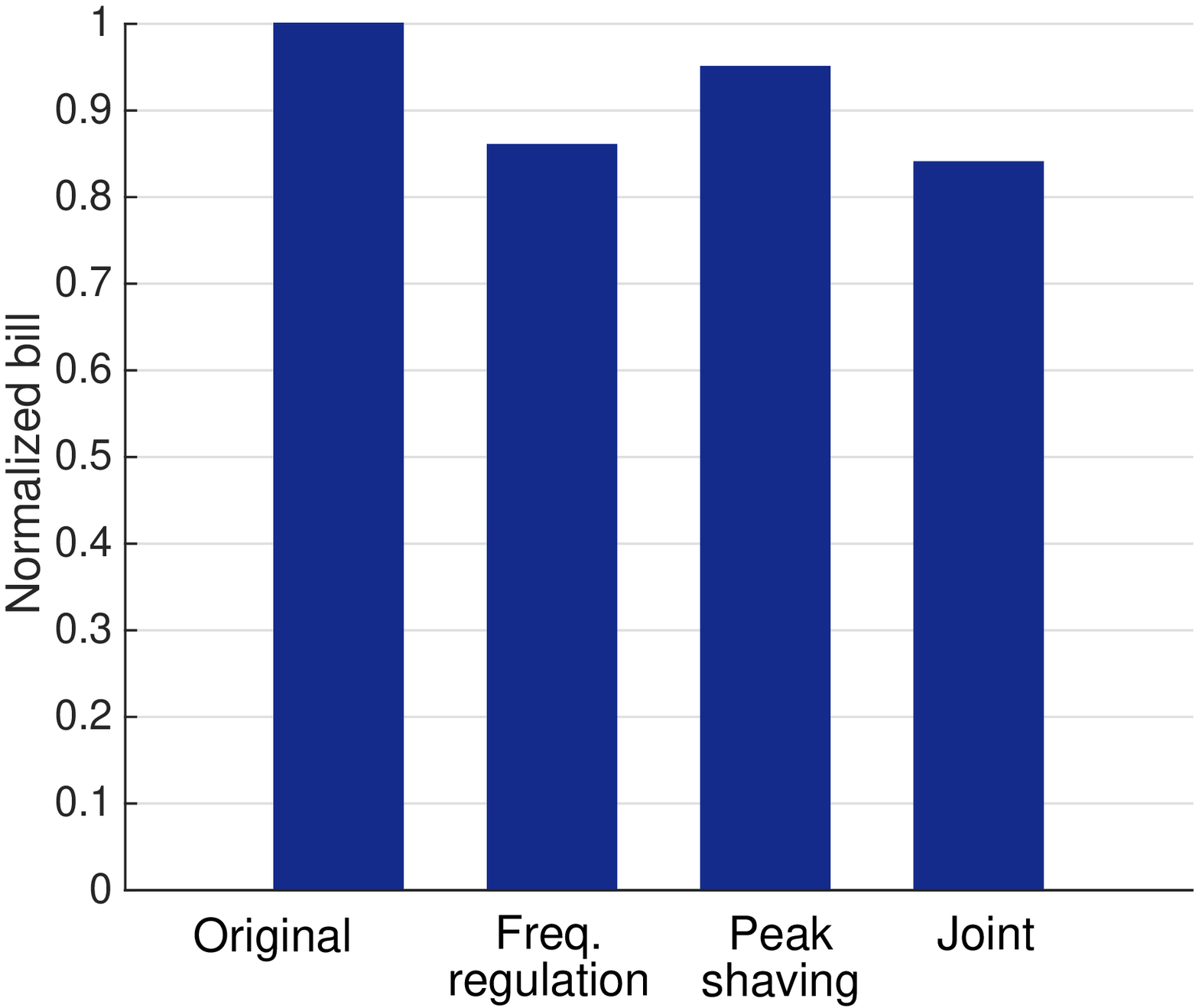}}
	\label{Sec4:P2:4}
	\caption{Electricity bills for a narrow peak (base load 0.5 MW, peak load 1MW, peak duration 3 minutes). \textbf{Labels:} In subfigures (a), (b), (c), the upper plot is power consumption; the lower plot is battery SoC curve. Blue solid line is the original load; red dotted dash line denotes demand+frequency regulation signal; green dashed line is the actual net consumption. Fig. d are normalized bills where the original bill is set to 1.}
	\label{Sec4:P2:s1}
\end{figure}
\begin{figure}
	\centering
	\subfloat[Frequency regulation only]{%
		\includegraphics[width=0.5\linewidth]{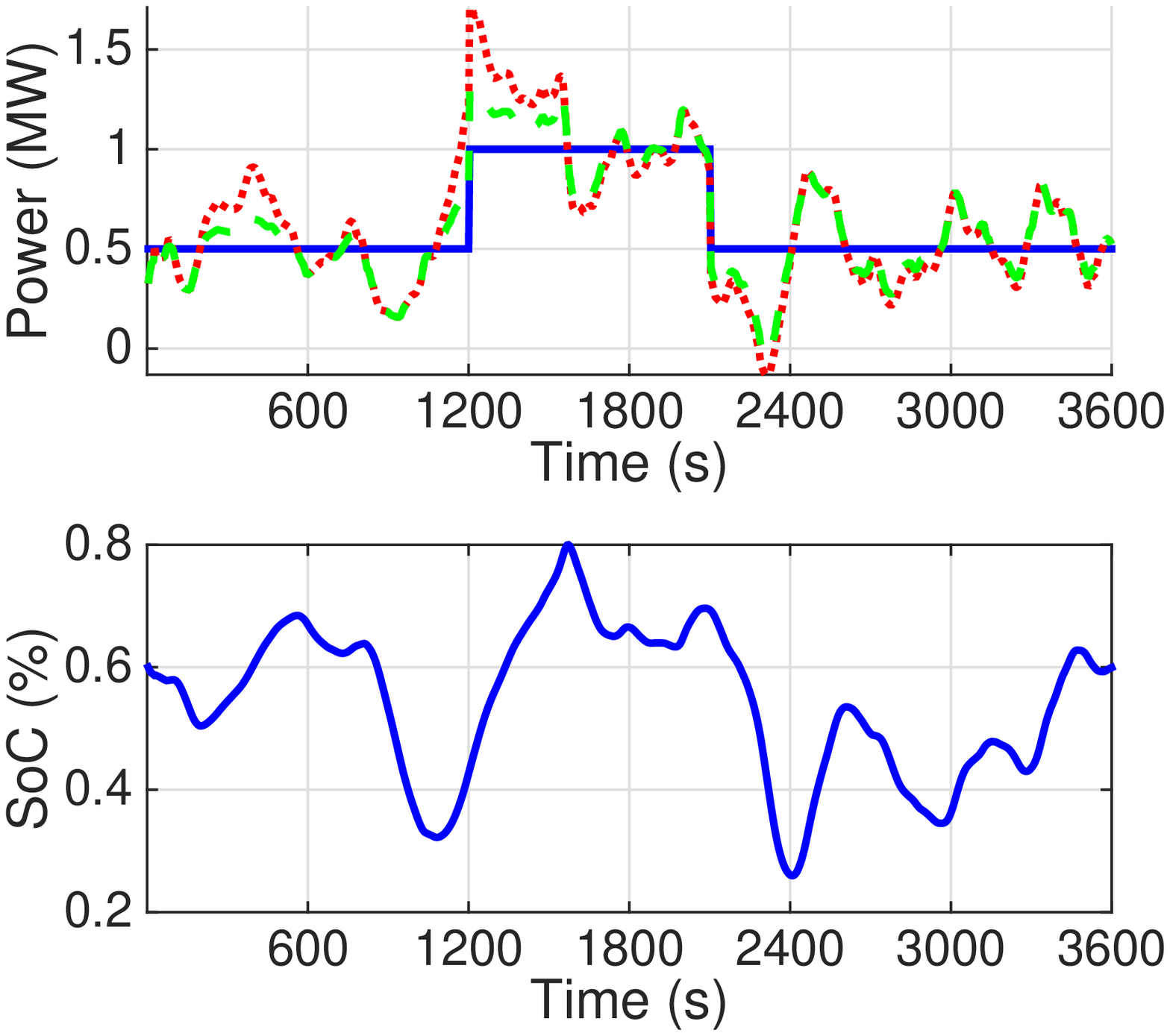}}
	\label{Sec4:P2:5}\hfill
	\subfloat[Peak shaving only]{%
		\includegraphics[width=0.5\linewidth]{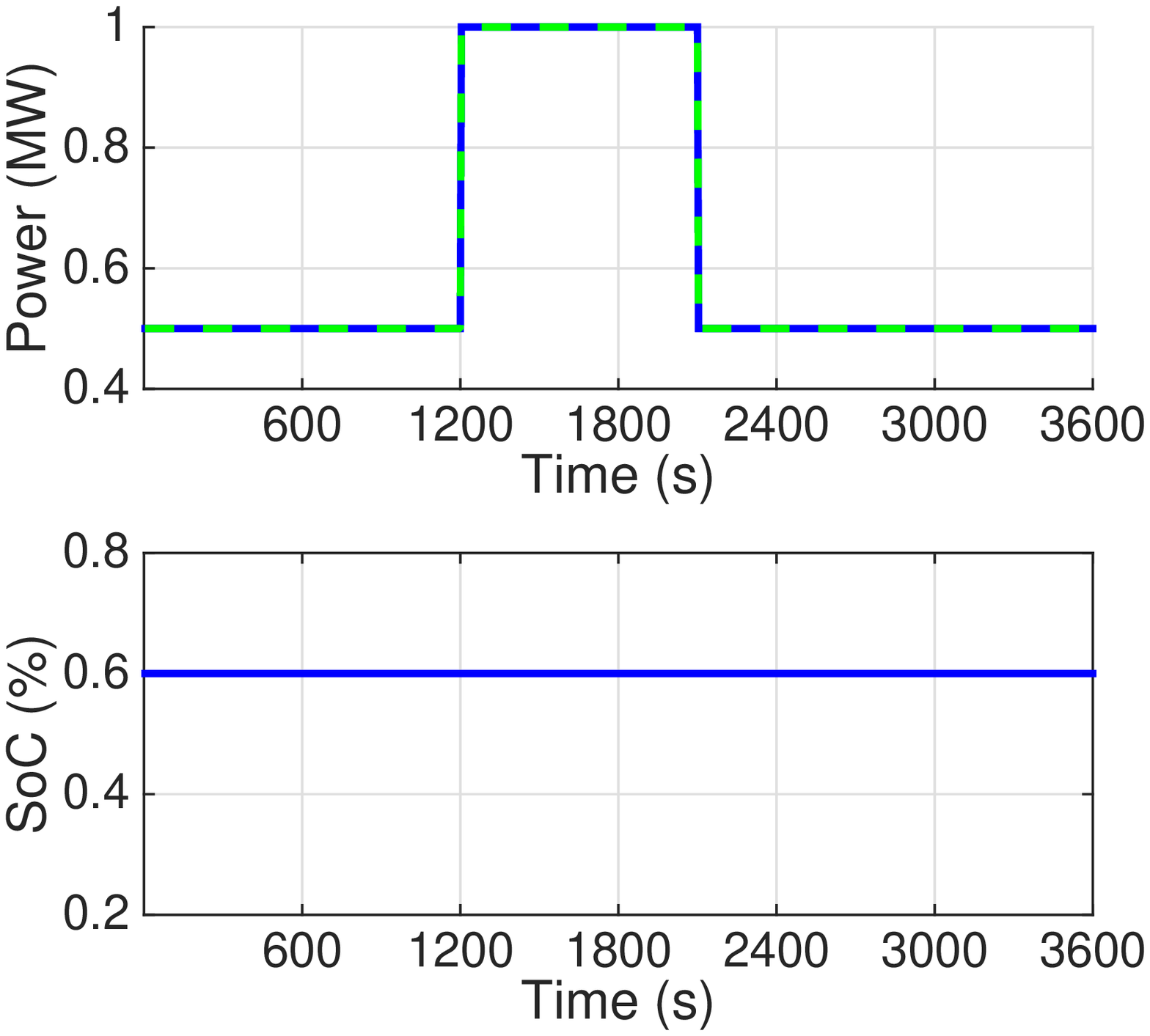}}
	\label{Sec4:P2:6}\\
	\subfloat[Joint optimization]{%
		\includegraphics[width=0.5\linewidth]{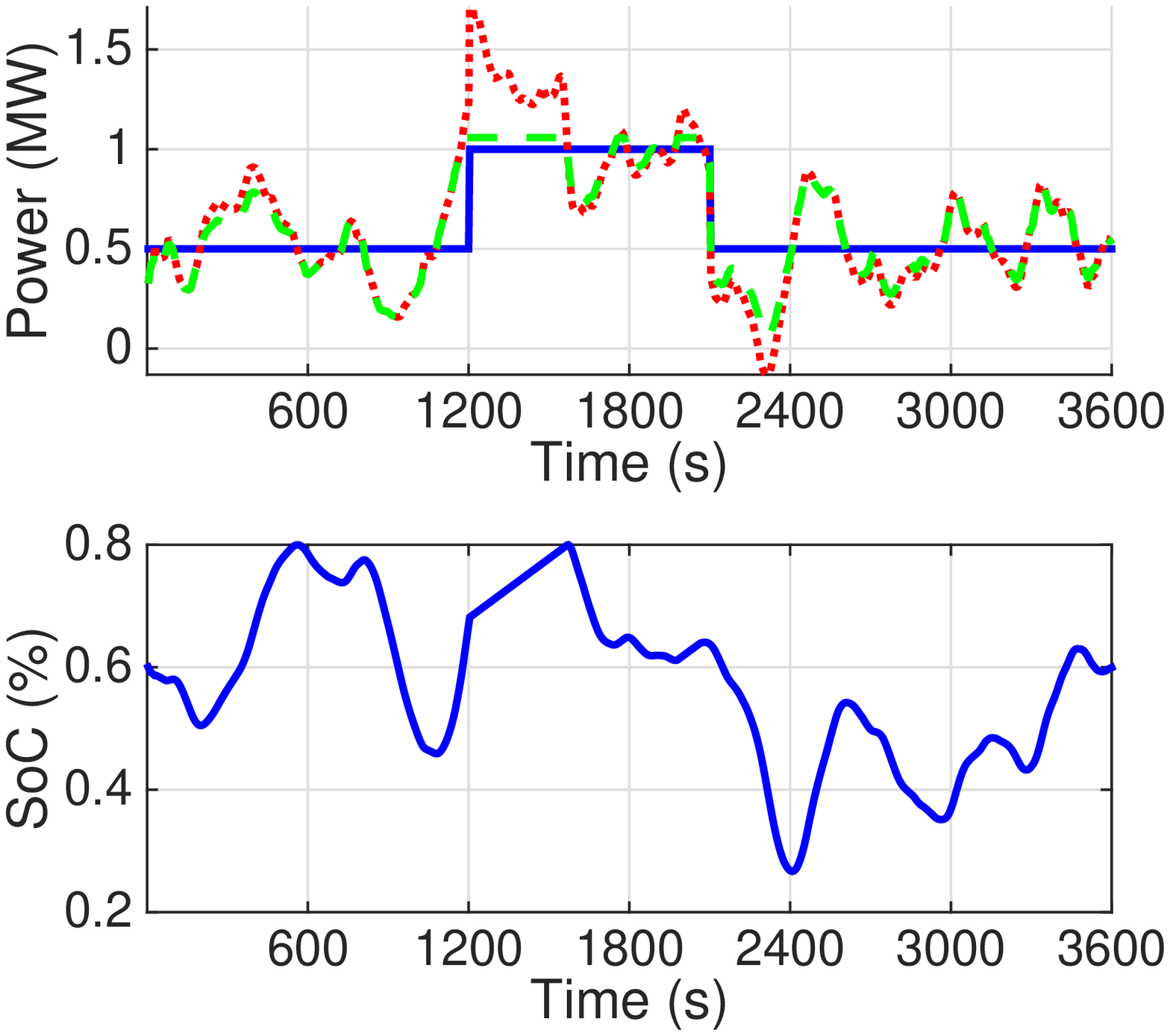}}
	\label{Sec4:P2:7}\hfill
	\subfloat[Bills comparison]{%
		\includegraphics[width=0.5\linewidth]{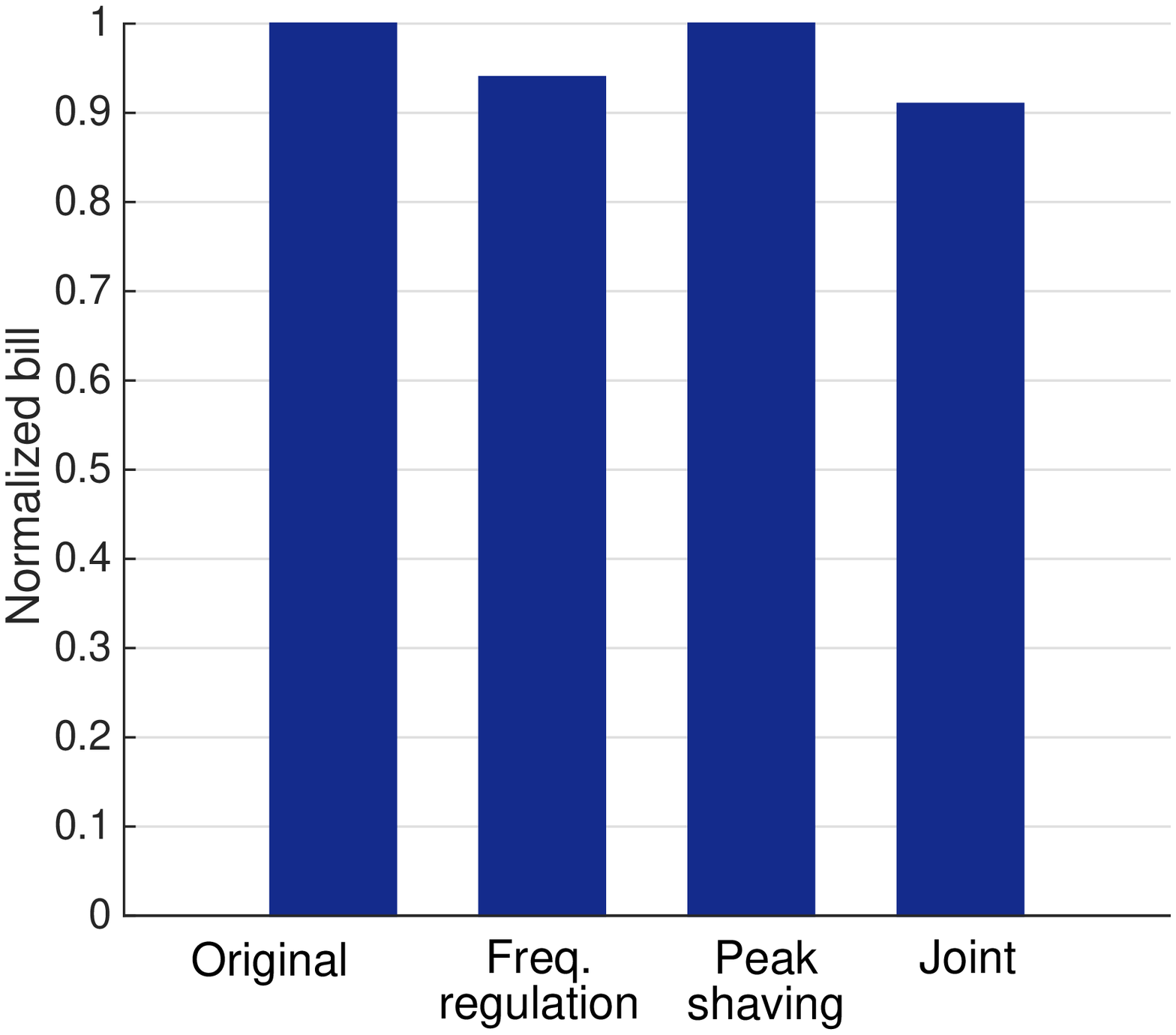}}
	\label{Sec4:P2:8}
	\caption{Electricity bills for a wide peak (base load 0.5 MW, peak load 1MW, peak duration 15 minutes). \textbf{Labels:} In subfigures (a), (b), (c), the upper plot is power consumption; the lower plot is battery SoC curve. Blue solid line is the original load; red dotted dash line denotes demand+frequency regulation signal; green dashed line is the actual net consumption. Fig. d are normalized bills where the original bill is set to 1.}
		\label{Sec4:P2:s2}
\end{figure}

\edit{Intuitively, the superlinear gain is related to the shape of demand curve. Consider two different peaks, a narrow peak (Fig. \ref{Sec4:P2:s1}) and a wide peak (Fig. \ref{Sec4:P2:s2}), we find that the main difference lies in the peak shaving part. For a 3 minute short-time peak, the battery could shave a large portion of the peak before hitting the SoC bound (Fig.\ref{Sec4:P2:s1} (b)). Thus, we save a lot from only doing peak shaving and the two applications do not interact much , and there is no superlinear saving. However, when the peak duration is long, it takes more battery energy to shave the same height off the peak. As seen from Fig.\ref{Sec4:P2:s2} (b), the battery doesn't respond much in the peak shaving only case because the cost of using battery gets close (or even exceeds) the saving from reduced peak demand charge. This argument is verified by Fig.\ref{Sec4:P2:s2} (d), where we find only doing peak shaving does not reduce the bill much. But if we consider joint optimization, the randomness of regulation signal helps break down the one flat peak into several short-time peaks, and we could save more from doing peak shaving on top of providing regulation service. This is where the superlinear saving comes from.}

\subsection{Results for Real-life Data: Microsoft Data Center and UW EE \& CSE Building}
This section conducts simulations based on real-life data from Microsoft data center and UW EE \& CSE building. Table \ref{Sec5:T1} and \ref{Sec5:T2} summarize the simulation results. We consider using a 1MW, 3 minutes battery for grid service, and the reported numerical results are achieved by the implementing the proposed simple online control algorithm. For a 1MW data center with \$488,370 annual electricity bill, the cost saving by joint optimization is around \$52,282 ($10.72\%$), with \$13,234 extra saving compared with the \emph{sum} of benchmark optima. For UW EE \& CSE building, 362 out of the 365 days, we have the superlinear gain. The annual electricity bill for UW EE \& CSE building is around \$359,634, from which we save \$44,420 ($12.35\%$) by implementing battery joint optimization. The superlinear gain is \$14,061 per year.
\begin{table}
   \renewcommand{\arraystretch}{1.5}
	\centering
	\caption{Half year simulation results of Microsoft data center}
	\begin{tabular}{ll}
		\hline
		\hline
	    \multicolumn{2}{c}{Microsoft data center} \\
		\hline
		Total days simulated & 183\\
		Days having superlinear gain & 151\\
		Average bill saving by joint optimization & 10.72\% \\
		Super linear saving ratio ($q$) & 2.71\% \\
		\hline
		\hline
	\end{tabular}
	\label{Sec5:T1}
\end{table}
\begin{table}
	\renewcommand{\arraystretch}{1.5}
	\centering
	\caption{One year simulation results of UW EE \& CSE building}
	\begin{tabular}{ll}
		\hline
		\hline
		\multicolumn{2}{c}{UW EE \&CSE building} \\
		\hline
		Total days simulated & 365 \\
		Days having superlinear gain & 362 \\
		Average bill saving by joint optimization &  12.35\%\\
		Super linear saving ratio ($q$) &  3.91\%\\
		\hline
		\hline
	\end{tabular}
	\label{Sec5:T2}
\end{table}

\begin{figure}
 	\centering
 	\subfloat[Microsoft data center]{%
 	\includegraphics[width= 0.4\columnwidth]{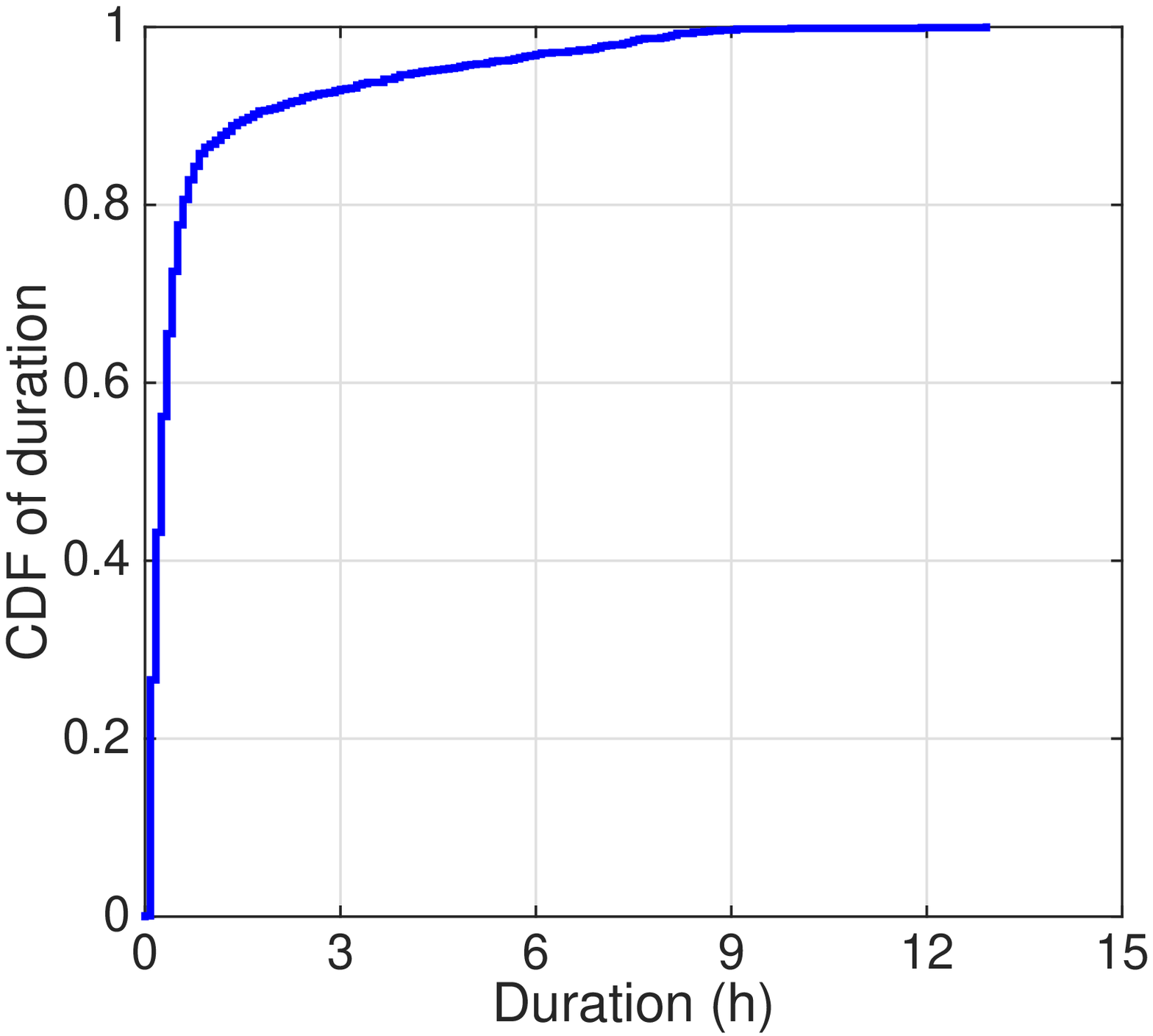}}
 	\label{fig:peakduration_dc}\hfill
 	\subfloat[UW EE \& CSE building]{%
 	\includegraphics[width=0.4\columnwidth]{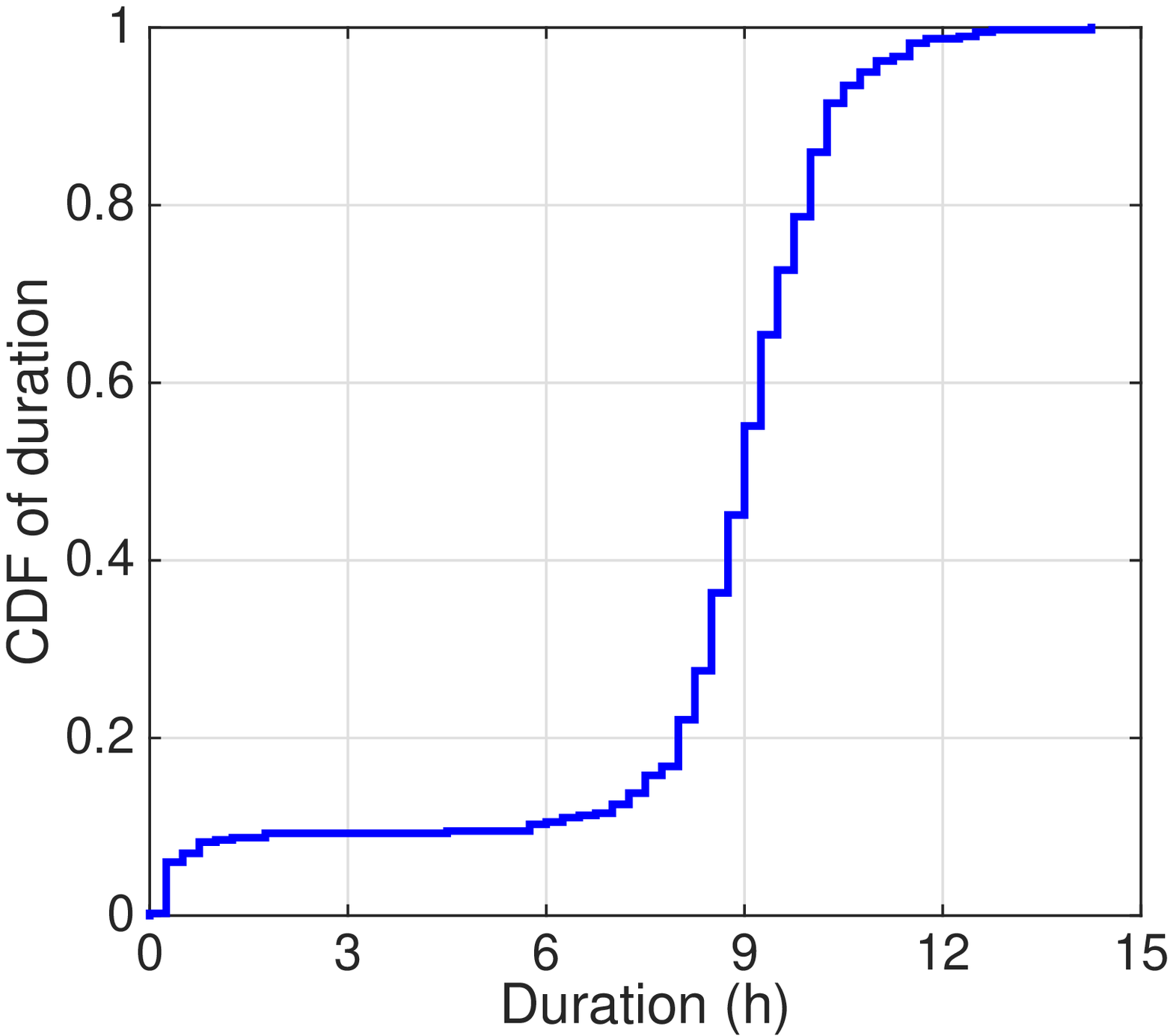}}
 	\label{fig:peakduration_eecs}\\
 	\caption{Cumulative distribution of peak duration for two case studies.}
 	\label{fig:stat_analysis}
 \end{figure}

 \edit{To link the analysis of synthetic load in the previous section to the real-life cases of data center and UW EE \& CSE building, we preform a statistical analysis of peak durations. We plot the Cumulative Distribution Function (CDF) of peak duration for the data center and UW EE \& CSE building in Fig. \ref{fig:stat_analysis}, where the average peak duration for data center is 0.75h (about 45 minutes) and 8.33h for the building.}

\edit{According to the observations in Figures 6 and 7, the proposed battery joint optimization has a larger gain for flat peaks compared to sharp peaks. Since the randomness of regulation signal helps break down one flat peak into several short-time peaks, we could save more from doing peak shaving on top of providing frequency regulation service. The exact definition of ``long'' and ``short'' peaks depend on the size of the battery. For a 3 minute battery, if the peak is shorter than 3 minutes, then performing joint optimization is not critical and we do not have a superlinear gain. On the other hand, for a peak that is longer than 3 minutes, it is important to use the regulation signal to break it up into smaller peaks. Therefore, both case studies have high superlinear gain probability, which is greater than 80\%. The superlinear gain ratio of UW building (99\%) is higher than the ratio of the data center (82.5\%) since there are virtually no peaks shorter than the battery capacity in the former and a still a few short peaks in the latter.}

\subsection{Sensitivity Analysis}
\begin{figure}
	\centering
	\includegraphics[width=0.6\columnwidth]{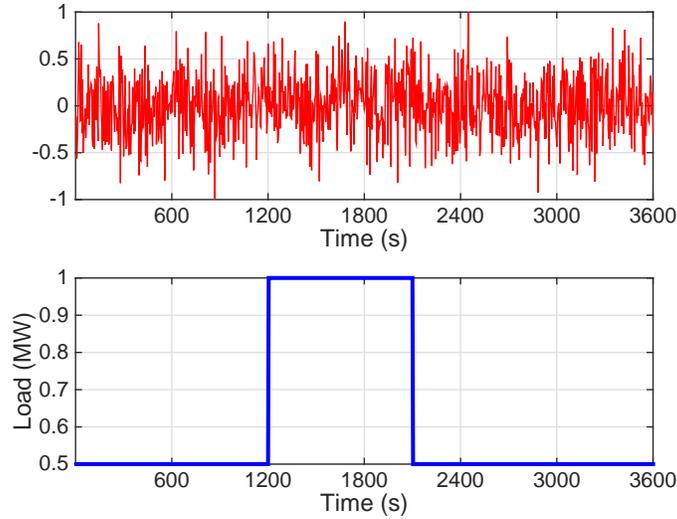}
	\caption{Synthetic frequency regulation signal (on the top) and rectangle peak (bottom) used for sensitivity analysis.}
	\label{fig:loadreg}
\end{figure}

Here, we preform sensitivity analysis about how different price settings, including different demand charge prices $\lambda_{peak}$, battery degradation costs $\lambda_{b}$ and regulation payments $\lambda_c$, influence the superlinear gain ratio. In order to quantatively evaluate the conditions when superlinear gain will happen and generalize the analysis to all potential scenarios, we pick a simple load curve with a rectangle peak (base load 0.5 MW, peak load is 1 MW, peak duration 15 minutes) and a truncated Gaussian signal as frequency regulation signal, with $\mu = 0$, $\sigma^{2} = 0.12$ (variance of the PJM RegD signal) and range $[-1,1]$. The simple rectangle peak and synthetic frequency regulation signal are given in Fig. \ref{fig:loadreg}.

\edit{Fig. 10a shows how the chance of having superlinear gain changes with regard to battery cell price and peak demand charge. The probability of superlinear gain increases as the battery cell price goes down, or as the peak demand charge goes up. The ``physical origin'' of the superlinear gain is the positive interaction between peak shaving and frequency regulation service. Since the randomness in the frequency regulation signal breaks the flat peak into several smaller peaks, more savings are obtained by performing peak shaving on the top of frequency regulation. Therefore, as the peak demand price goes up, it yields more economic benefits to jointly optimize the two applications. Similarly, the as the battery prices decreases, it can be used more aggressively for both applications. As battery prices continue to decrease in the future, the benefits of joint optimization will increase.}

\edit{Fig. 10b demonstrates how the probability of having superlinear relates to battery cell price and regulation capacity payment. The chance of having superlinear gain is the highest when both the battery cell price and regulation capacity payment are low. When the capacity payment is high enough, it yields much more economic benefits to provide frequency regulation service than peak shaving. In such condition, the  probability of having superlinear gain decreases.}
\begin{figure}
	\centering
	\subfloat[Probability of superlinear gain V.S. battery cell price $\lambda_{cell}$(\$/Wh) and peak demand charge $\lambda_{peak}$(\$/MW)]{%
	\includegraphics[width= 0.4\columnwidth]{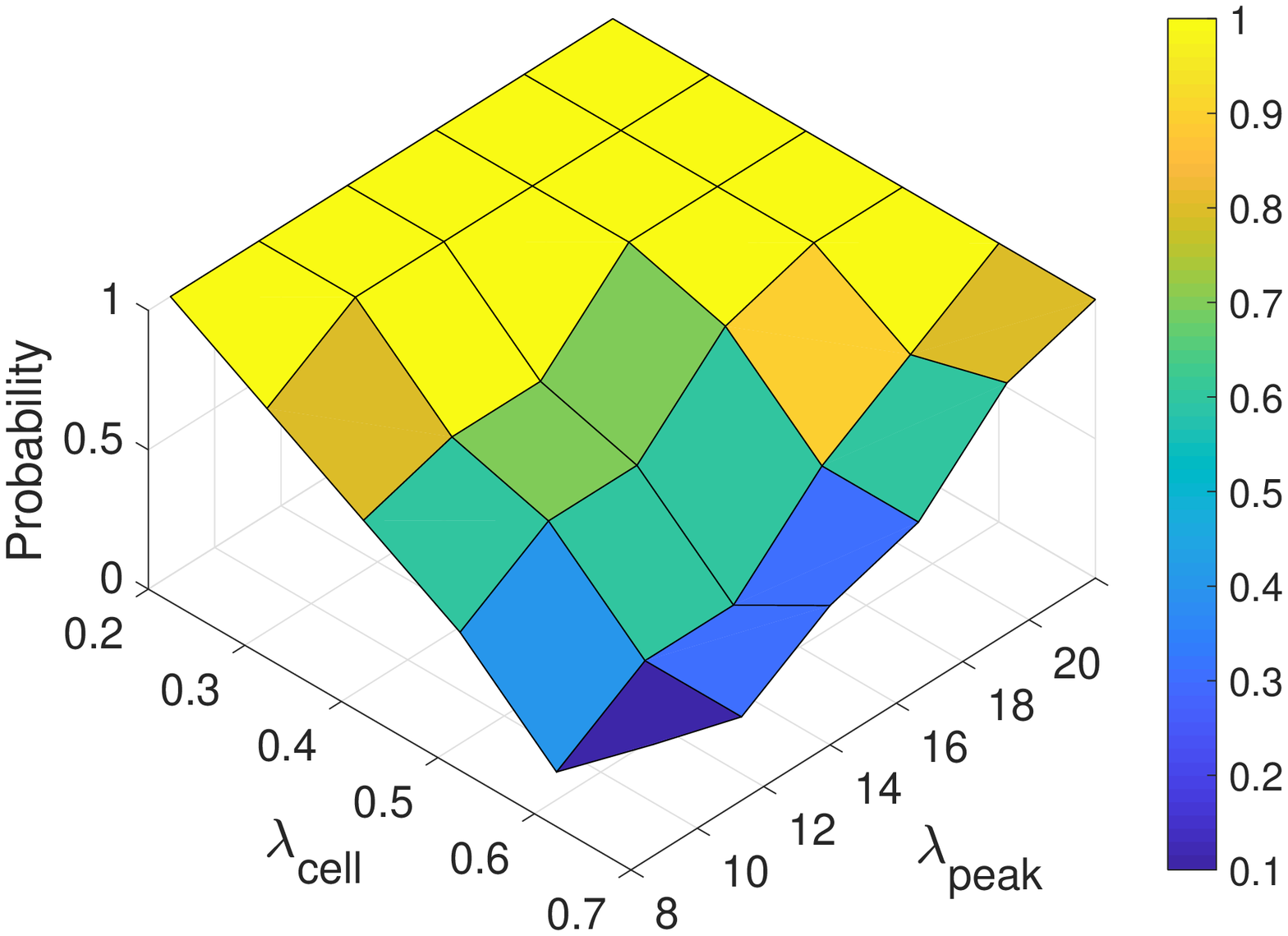}}
	\label{fig:analysis_cellpeak}
	\subfloat[Probability of superlinear gain V.S. battery cell price $\lambda_{cell}$(\$/Wh) and regulation capacity payment $\lambda_c$(\$/MWh)]{%
	\includegraphics[width=0.4\columnwidth]{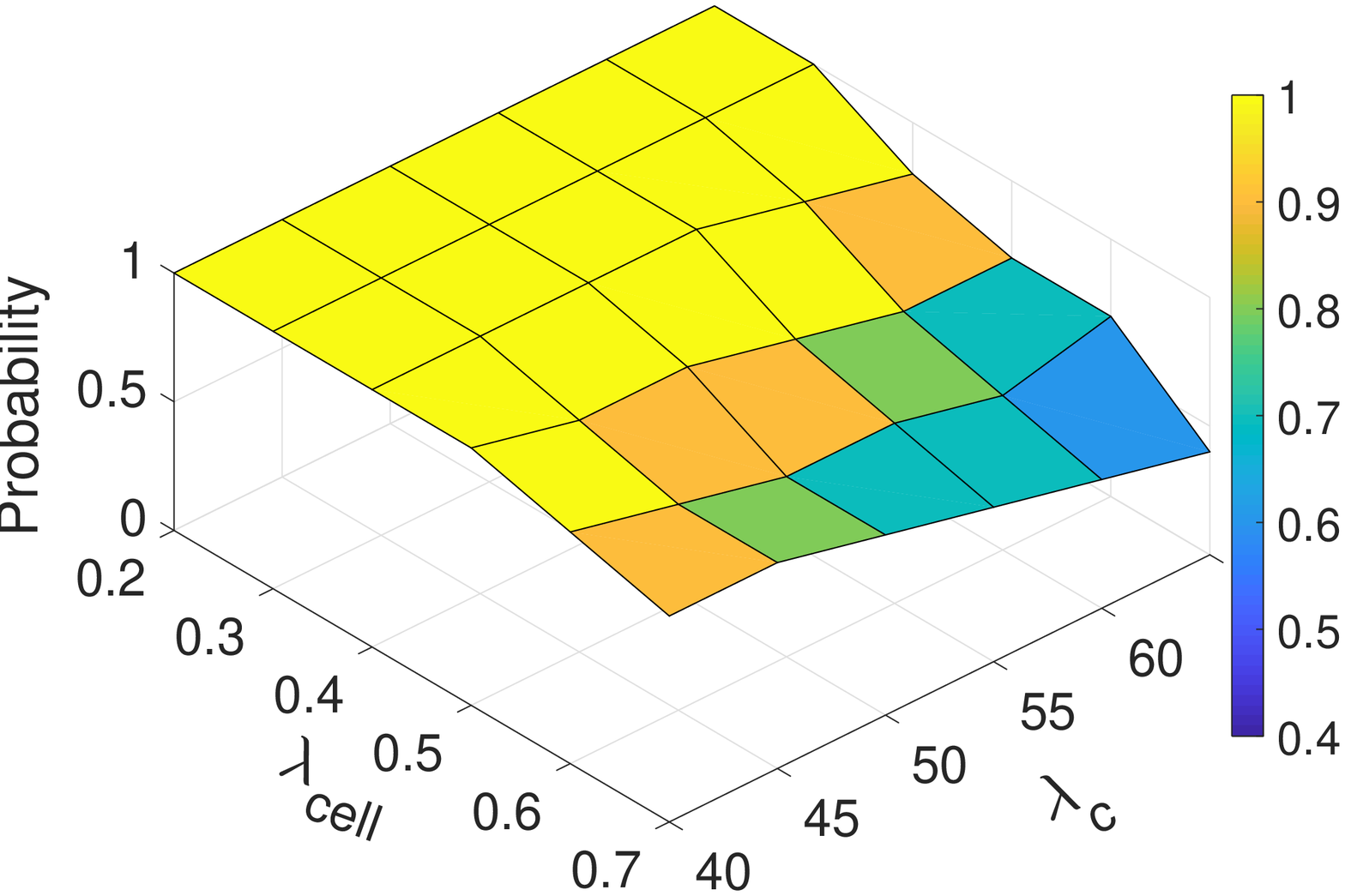}}
	\label{fig:analysis_cellregc}\\
	\caption{Superlinear gain probability V.S. price coefficients}
	\label{fig:analysis_price}
\end{figure}

\section{Conclusion}
\label{sec:con}
This paper addresses using battery storage in large commercial users to reduce their electricity bills. We consider two sources of cost savings: reducing the peak demand charge and gaining revenue from participating in frequency regulation market. We formulate a framework that jointly optimizes battery usage for both of these applications. Surprisingly, we observe that a superlinear gain can often be obtained: the savings from joint optimization can be larger than the sum of the individual savings from devoting the battery to one of the applications. We also developed an online control algorithm which achieves the superlinear gain. The battery degradation model presented in this paper is in simplified linear form, which applies to certain battery operation range. Incorporating a more general and accurate battery degradation model, such as cycle-based degradation model~\cite{shi2017convex} into the joint optimization framework might be an interesting direction for future work.

\appendices
\section{Commertial User's Load Prediction}
\label{sec:appden1}
Solving the stochastic joint optimization problem in (\ref{Sec3:P4:1}) requires accurate short-term load forecasting (STLF) for the next 24 hours. A lot of research has been done in the area of STLF. There are two major factors determine the quality of load prediction, \emph{input features} and prediction \emph{model}. On the one hand, selecting features or a group of features which affect the future load most is important. The input features mainly include the effect of nature (eg. temperature) and the effects of human activities (calendar variables, e.g., business hours), and the interaction of above two factors. On the other hand, deciding which kind of models to forecast future load is also crucial. People have been adopting or developing various techniques for day-ahead load forecasting, including regression, time series analysis, neural networks, support vector machine and a combination of the above methods (see \cite{hong2010} and references within).

\begin{figure}
	\centering
	\includegraphics[width=0.5\columnwidth]{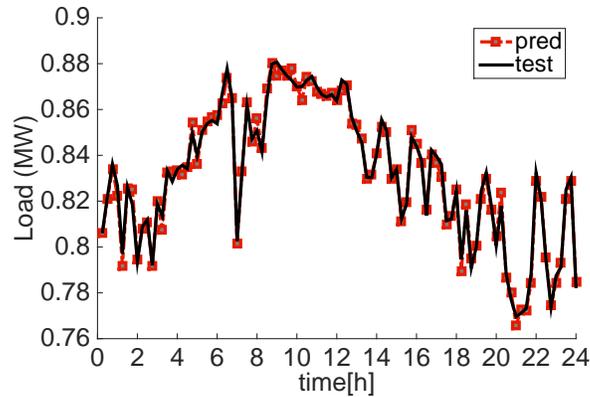}
	\caption{Load prediction for data center demand, the black curve is the actual demand, and the red line are the day ahead load prediction using MLR. The load is scaled between 0 and 1MW.}
	\label{fig:pred_datacenter}
\end{figure}

In this paper, we used a multiple linear regression (MLR) model that takes $\mathcal{X}=\{$trend, temperature forecasting (TMP), month, Hour $\times$ TMP, month $\times$ TMP, day $\times$ Hour, adjacency day's load, weekend and holiday effect, recent similar days' average$\}$ as input, and use the following MLR model to predict the power demand for next 1 day. Fig \ref{fig:pred_datacenter} presents the day-ahead load prediction result for a data center.

\begin{align}\label{equ:mlr_pred}
Y & = \beta_{0} + \beta_1 \times Trend + \beta_2 \times TMP + \beta_3 \times Month \nonumber\\
& +\beta_4 \times Hour \times TMP + \beta_5 \times month \times TMP \nonumber\\
& + \beta_6 \times day \times Hour + \beta_7 \times Load(day-1) \\
& + \beta_8 \times weekend +\beta_9 \times holiday + \beta_{10} \times \bar{Load}(day-1) \nonumber
\end{align}

\section{Frequency regulation signal scenarios reduction}
\label{sec:appden2}
In order to solve the stochastic joint optimization problem in (\ref{Sec3:P4:1}), we also need to model the uncertainty of future regulation signals. In this paper, we use one-year historical data to empirically model the distributions of regulation signals. Each daily realization of the regulation signal is called a ``scenario'', and thus we obtain 365 scenarios.

\begin{figure}
	\centering
	\includegraphics[width=0.5\columnwidth]{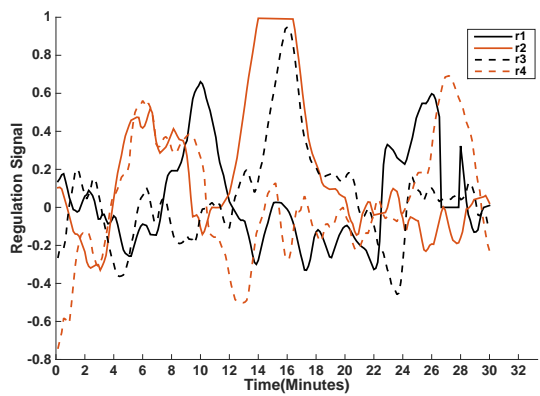}
	\caption{Selected frequency regulation signal scenarios. r1, r2, r3, r4 are the top four representative frequency regulation signal scenarios.}
	\label{fig:scenario}
\end{figure}

Because a large number of scenarios will reduce the computational tractability of the joint optimization problem, it is useful to choose a smaller subset of scenarios that can well approximate the original entire scenario set. We applied the forward scenario reduction algorithm in~\cite{wang2015} to select the best subset of scenarios, and assign new probabilities to the selected scenarios. The key idea of scenario reduction is to pick a subset of scenarios which preserve as much information as the original set. We set the number of selected scenarios as 10, which strives for a balance between performance and computational complexity by simulation. For visualization clarity, we plot 4 out of the 10 selected scenarios in Fig. \ref{fig:scenario}.

\section{Proof of Theorem 1}
\label{sec:appden3}
Here we provide a detailed proof of Theorem 1 in \ref{sec:sim_control}.
\begin{proof}
	Under linear battery cost model, it is obvious that $b^*(t)$ and $r(t)$ always have the same sign. Or equivalent saying,  $b(t)$ and $r(t)$ are always both positive or both negative. Based on the relative sizes of coefficients and sign, there are 5 cases to be considered:
	\begin{equation*}
	\begin{cases}
	\lambda_b < \lambda_{mis}
	\begin{cases}
	r(t) \geq 0
	\begin{cases}
	b(t) \geq Cr(t) \ (i)\\
	b(t) < Cr(t) \ (ii)
	\end{cases}\\
	r(t) <0
	\begin{cases}
	b(t) \geq Cr(t) \ (iii)\\
	b(t) < Cr(t) \ (iv)
	\end{cases}\\
	\end{cases}\\

	\lambda_b \geq \lambda_{mis} \ (v)
	\end{cases}
	\end{equation*}

	\textbf{(i).} $\lambda_b < \lambda_{mis}$, $r(t) \geq 0$ and $b(t) \geq Cr(t)$

	In this case, $b(t) \geq 0$, battery is discharging. And the objective function (\ref{reg:obj}) becomes,

	$$\underset{C,b(t)}{\text{maximize\ \ }} \lambda_c C + \frac{1}{T}E \{-(\lambda_{mis}+\lambda_b)\sum_{t=1}^{T}b(t)+\lambda_{mis}\sum_{t=1}^{T} Cr(t)\}$$\,,

	Notice that the coefficient in front of $b(t)$ is negative. In order to maximize the objective, we need to minimize $b(t)$ under the following constraints:
	\begin{equation*}
	\begin{cases}
	b(t) \geq 0\\
	b(t) \geq Cr(t) \\
	b(t) \leq P^{max} \\
	b(t) \leq \frac{\eta_d [soc(t)-SoC_{min}]E}{t_s}
	\end{cases}
	\end{equation*}

	\begin{itemize}
		\item If either $Cr(t) > P^{max}$ or $Cr(t) > \frac{\eta_d [soc(t)-SoC_{min}]E}{t_s}$, there is no feasible solution for $b(t)$.

		\item If $Cr(t) \leq P^{max}$ and $Cr(t) \leq \frac{\eta_d [soc(t)-SoC_{min}]E}{t_s}$, we have the optimal $b^*(t)=Cr(t)$.
	\end{itemize}

	\textbf{(ii).} $\lambda_b < \lambda_{mis}$, $r(t) \geq 0$ and $b(t) < Cr(t)$

	In this case, $b(t) \geq 0$, battery is discharging, and the objective function (\ref{reg:obj}) becomes,

	$$\underset{C,b(t)}{\text{maximize\ \ }} \lambda_c C + \frac{1}{T}E \{(\lambda_{mis}-\lambda_b)\sum_{t=1}^{T}b(t)-\lambda_{mis}\sum_{t=1}^{T} Cr(t)\}$$\,,

	Notice the coefficient in front of $b(t)$ is $(\lambda_{mis}-\lambda_b)$, which is positive in this case. So in order to maximize the objective, we need to maximize $b(t)$ under the following constraints:
	\begin{equation*}
	\begin{cases}
	b(t) \geq 0\\
	b(t) < Cr(t) \\
	b(t) \leq P^{max}\\
	b(t) \leq \frac{\eta_d [soc(t)-SoC_{min}]E}{t_s}
	\end{cases}
	\end{equation*}

	So $b^*(t) = min\{Cr(t),P^{max}, \frac{\eta_d [soc(t)-SoC_{min}]E}{t_s}\}$.

	Summarizing case (i, ii), we get if $\lambda_b < \lambda_{mis}$, $r(t) \geq 0$
	$$b^*(t) = min\{Cr(t),P^{max}, \frac{\eta_d [soc(t)-SoC_{min}]E}{t_s}\}\,.$$

	\textbf{(iii).} $\lambda_b < \lambda_{mis}$, $r(t) < 0$ and $b(t) \geq Cr(t)$

	In this case, $b(t) < 0$, battery is charging, and the objective function (\ref{reg:obj}) becomes,

	$$\underset{C,b(t)}{\text{maximize\ \ }} \lambda_c C + \frac{1}{T}E \{(-\lambda_{mis}+\lambda_b)\sum_{t=1}^{T}b(t)+\lambda_{mis}\sum_{t=1}^{T} Cr(t)\}$$\,,

	The coefficient in front of $b(t)$ is $(-\lambda_{mis}+\lambda_b)$, which is negative. So in order to maximize the objective, we need to minimize $b(t)$ under the following constraints:
	\begin{equation*}
	\begin{cases}
	b(t) \leq 0\\
	b(t) \geq Cr(t) \\
	b(t) \geq -P^{max} \\
	b(t) \geq \frac{[SoC(t)-SoC_{max}] E}{\eta_c t_s}
	\end{cases}
	\end{equation*}

	The minimal $b(t)$ is optimal, $b^*(t) = max\{Cr(t),-P^{max}, \frac{[SoC(t)-SoC_{max}] E}{\eta_c t_s}\}$.

	\textbf{(iv).} $\lambda_b < \lambda_{mis}$, $r(t) < 0$ and $b(t) < Cr(t)$

	In this case, $b(t) < 0$, battery is charging, and the objective function (\ref{reg:obj}) becomes,

	$$\underset{C,b(t)}{\text{maximize\ \ }} \lambda_c C + \frac{1}{T}E \{(\lambda_{mis}+\lambda_b)\sum_{t=1}^{T}b(t)-\lambda_{mis}\sum_{t=1}^{T} Cr(t)\}$$\,,

	The coefficient in front of $b(t)$ is positive, so in order to maximize the objective, we need to maximize $b(t)$ under the following constraints:
	\begin{equation*}
	\begin{cases}
	b(t) < 0\\
	b(t) < Cr(t) \\
	b(t) \geq -P^{max} \\
	b(t) \geq \frac{[SoC(t)-SoC_{max}] E}{\eta_c t_s}
	\end{cases}
	\end{equation*}

	\begin{itemize}
		\item If either $Cr(t) <  -P^{max}$ or $Cr(t) < \frac{[SoC(t)-SoC_{max}] E}{\eta_c t_s}$, there is no feasible solution for $b(t)$.

		\item If $Cr(t) \geq -P^{max}$ and $Cr(t) \geq \frac{[SoC(t)-SoC_{max}] E}{\eta_c t_s}$, we have the optimal $b^*(t)=Cr(t)$.
	\end{itemize}

	Summarizing case (iii, iv), we get if $\lambda_b < \lambda_{mis}$, $r(t) < 0$,
	$$b^*(t) = max\{Cr(t),-P^{max}, \frac{[SoC(t)-SoC_{max}] E}{\eta_c t_s}\}\,.$$

	\textbf{(v).} $\lambda_b \geq \lambda_{mis}$

	We know that $b(t)$ and $r(t)$ always have the same sign, so the objective function   (\ref{reg:obj}) could be expressed as,

	\begin{equation*}
	\lambda_c \cdot C - \frac{1}{T} E\{\sum_{t=1}^{T} \lambda_{mis} ||b(t)|-C|r(t)|| + \sum_{t=1}^{T} \lambda_b |b(t)|\}\,,
	\end{equation*}

	For each time step $t$, we take derivative of the objective function w.r.t. $|b(t)|$,
	\begin{equation*}
	\begin{aligned}
	\frac{\delta J}{\delta |b(t)|} & = - (\lambda_{mis} \cdot \vec{1}_{|b(t)>C|r(t)||}
	-  \lambda_{mis} \cdot \vec{1}_{|b(t)<C|r(t)||} + \lambda_b)\\
	& = - \lambda_{mis} \cdot \vec{1}_{|b(t)>C|r(t)||}
	+  \lambda_{mis} \cdot \vec{1}_{|b(t)<C|r(t)||} - \lambda_b\\
	& \leq \lambda_{mis} \cdot \vec{1}_{|b(t)<C|r(t)||} - \lambda_b\\
	& \leq \lambda_{mis} - \lambda_b \leq 0
	\end{aligned}
	\end{equation*}

	Since $\frac{\delta J}{\delta |b(t)|} \leq 0$, so in order to maximize $J$ (the regulation service benefits), $|b|=0$. Therefore, 	when $\lambda_b \geq \lambda_{mis}$, $b^*(t)=0$ is optimal for $\forall C \geq 0$.
\end{proof}

\bibliographystyle{ieeetr}
\bibliography{sigproc,dc_bib}  

\begin{thebibliography}{10}

\bibitem{CAISO2014}
{CAISO}, ``Advancing and maximizing the value of energy storage technology.''
  Technical Report, 2014.

\bibitem{SolomonEtAl2014}
A.~Solomon, D.~M. Kammen, and D.~Callaway, ``The role of large-scale energy
  storage design and dispatch in the power grid: a study of very high grid
  penetration of variable renewable resources,'' {\em Applied Energy},
  vol.~134, pp.~75--89, 2014.

\bibitem{WorthmannEtAl2015}
K.~Worthmann, C.~M. Kellett, P.~Braun, L.~Gr{\"u}ne, and S.~R. Weller,
  ``Distributed and decentralized control of residential energy systems
  incorporating battery storage,'' {\em IEEE Transactions on Smart Grid},
  vol.~6, no.~4, pp.~1914--1923, 2015.

\bibitem{Guo2013}
Y.~Guo and Y.~Fang, ``Electricity cost saving strategy in data centers by using
  energy storage,'' {\em IEEE Transactions on Parallel and Distributed
  Systems}, vol.~24, pp.~1149--1160, June 2013.

\bibitem{WangEtAl2016}
Y.~Wang, B.~Wang, C.-C. Chu, H.~Pota, and R.~Gadh, ``Energy management for a
  commercial building microgrid with stationary and mobile battery storage,''
  {\em Energy and Buildings}, vol.~116, pp.~141--150, 2016.

\bibitem{CVD2014}
{Cisco Validated Designs}, ``Data center technolog design guide,'' tech. rep.,
  Cisco, 2014.

\bibitem{EPRI2013}
{EPRI}, ``Grid energy storage,'' tech. rep., Department of Energy, 2013.

\bibitem{Shi2016}
Y.~Shi, B.~Xu, B.~Zhang, and D.~Wang, ``Leveraging energy storage to optimize
  data center electricity cost in emerging power markets,'' in {\em Proceedings
  of the Seventh International Conference on Future Energy Systems}, e-Energy
  '16, pp.~18:1--18:13, ACM, 2016.

\bibitem{DunnEtAl2011}
B.~Dunn, H.~Kamath, and J.-M. Tarascon, ``Electrical energy storage for the
  grid: a battery of choices,'' {\em Science}, vol.~334, no.~6058,
  pp.~928--935, 2011.

\bibitem{eyer2010energy}
J.~Eyer and G.~Corey, ``Energy storage for the electricity grid: Benefits and
  market potential assessment guide,'' {\em Sandia National Laboratories},
  vol.~20, no.~10, p.~5, 2010.

\bibitem{Wasowicz2012}
B.~Wasowicz, S.~Koopmann, T.~Dederichs, A.~Schnettler, and U.~Spaetling,
  ``Evaluating regulatory and market frameworks for energy storage deployment
  in electricity grids with high renewable energy penetration,'' in {\em In
  European Energy Market (EEM), 2012 9th International Conference on the ,
  IEEE}, 2012.

\bibitem{xi2014stochastic}
X.~Xi, R.~Sioshansi, and V.~Marano, ``A stochastic dynamic programming model
  for co-optimization of distributed energy storage,'' {\em Energy Systems},
  vol.~5, no.~3, pp.~475--505, 2014.

\bibitem{cheng2016co}
B.~Cheng and W.~Powell, ``Co-optimizing battery storage for the frequency
  regulation and energy arbitrage using multi-scale dynamic programming,'' {\em
  IEEE Transactions on Smart Grid}, 2016.

\bibitem{walawalkar2007economics}
R.~Walawalkar, J.~Apt, and R.~Mancini, ``Economics of electric energy storage
  for energy arbitrage and regulation in new york,'' {\em Energy Policy},
  vol.~35, no.~4, pp.~2558--2568, 2007.

\bibitem{white2011using}
C.~D. White and K.~M. Zhang, ``Using vehicle-to-grid technology for frequency
  regulation and peak-load reduction,'' {\em Journal of Power Sources},
  vol.~196, no.~8, pp.~3972--3980, 2011.

\bibitem{dowling2017multi}
A.~W. Dowling, R.~Kumar, and V.~M. Zavala, ``A multi-scale optimization
  framework for electricity market participation,'' {\em Applied Energy},
  vol.~190, pp.~147--164, 2017.

\bibitem{Sigrist2013}
S.~Lukas, E.~Lobato, and L.~Rouco, ``Energy storage systems providing primary
  reserve and peak shaving in small isolated power systems: an economic
  assessment,'' {\em International Journal of Electrical Power \& Energy
  Systems}, vol.~53, pp.~675--683, December 2013.

\bibitem{Caprino2014}
D.~Caprino, M.~L.~D. Vedova, and T.~Facchinetti, ``Peak shaving through
  real-time scheduling of household appliances,'' {\em Energy and Buildings},
  vol.~75, pp.~133--148, June 2014.

\bibitem{xu2016pesgm}
B.~Xu, Y.~Dvorkin, D.~S. Kirschen, C.~A. Silva-Monroy, and J.~P. Watson, ``A
  comparison of policies on the participation of storage in us frequency
  regulation markets,'' in {\em In Power and Energy Society General Meeting
  (PESGM)}, 2016.

\bibitem{pjm}
``Pjm historical regulation market data.''
\newblock Available online
  \url{http://www.pjm.com/markets-and-operations/ancillary-services.aspx}.

\bibitem{Miguel2011}
O.~V.~M. A., ``Optimal scheduling of electric vehicle charging and
  vehicle-to-grid services at household level including battery degradation and
  price uncertainty,'' {\em IET Generation, Transmission \& Distribution},
  vol.~8, pp.~1007--1016, January 2011.

\bibitem{Xu2016degradation}
B.~Xu, A.~Oudalov, A.~Ulbig, G.~Andersson, and D.~Kirschen, ``Modeling of
  lithium-ion battery degradation for cell life assessment,'' {\em IEEE
  Transactions on Smart Grid}, June 2016.

\bibitem{cvx}
M.~Grant and S.~Boyd, ``Cvx: Matlab software for disciplined convex
  programming, version 2.0 beta,'' 2013.

\bibitem{hong2010}
H.~Tao, {\em Short term electric load forecasting}.
\newblock North Carolina State University, 2010.

\bibitem{wang2015}
H.~Wang and J.~Huang, ``Cooperative planning of renewable generations for
  interconnected microgrids,'' {\em IEEE Transactions on Smart Grid}, vol.~7,
  pp.~2486--2496, December 2016.

\bibitem{qin2016}
J.~Qin, Y.~Chow, J.~Yang, and R.~Rajagopal, ``Online modified greedy algorithm
  for storage control under uncertainty,'' {\em IEEE Transactions on Power
  Systems}, vol.~31, pp.~1729--1743, May 2016.

\bibitem{wang2012energy}
D.~Wang, C.~Ren, A.~Sivasubramaniam, B.~Urgaonkar, and H.~Fathy, ``Energy
  storage in datacenters: what, where, and how much?,'' in {\em ACM SIGMETRICS
  Performance Evaluation Review}, vol.~40, pp.~187--198, ACM, 2012.

\bibitem{shi2017convex}
Y.~Shi, B.~Xu, Y.~Tan, and B.~Zhang, ``A convex cycle-based degradation model
  for battery energy storage planning and operation,'' {\em arXiv preprint},
  2017.

\end{thebibliography}
\end{document}